# The Jupiter ICy moons Explorer (JUICE) Mission: Post-Launch Status and Mission Overview

**Olivier Witasse · Giuseppe Sarri · Christian Erd · Philippe Gare · Nicolas Altobelli · Claire Vallat et al.** ***[full author details at the end of the article]***

**Abstract**
The JUpiter ICy Moons Explorer (JUICE) is the first large-class mission in ESA's Cosmic Vision programme. Proposed in 2007 and selected in 2012, it was launched on 14 April 2023 from Kourou and is now *en route* towards Jupiter. After arrival in July 2031, JUICE will spend approximately four years studying the Jupiter system, performing multiple flybys of three of its Galilean moons, the icy moons Ganymede, Europa, and Callisto, before entering orbit around Ganymede for its final year. The mission will investigate conditions that may support habitable environments within the satellites, all of which are thought to contain subsurface liquid water oceans. Ganymede, the largest moon in the Solar System, is a prime target due to its internal structure and possible subsurface ocean, its unique intrinsic magnetic field, and its complex interactions with Jupiter's magnetosphere, offering a natural laboratory for studying the emergence, evolution and potential habitability of icy worlds around giant planets. JUICE will also conduct a global investigation of the Jupiter system as the archetype for gaseous giant planet systems: it will study Jupiter's atmospheric circulation, composition, and flow of energy from the cloud tops to the thermosphere and characterise the 3D structure of the magnetosphere and its coupling with the ionosphere and thermosphere. Interactions between the moons and the magnetosphere, as well as tidal and gravitational evolution within the Galilean system, will also be assessed. Additional science objectives include the innermost Galilean moon, Io and its neutral cloud and plasma torus, some small inner and irregular moons, and the ring systems. The payload includes four remote-sensing instruments, three geophysical packages, three particles-and-fields instruments, and one experiment using the spacecraft's telecommunications system to connect with ground-based radio telescopes, plus an onboard radiation monitor.

## Abbreviations

| | |
|---|---|
| 3GM | Gravity and Geophysics of Jupiter and Galilean Moons |
| ADS | Airbus Defense and Space |
| AMDA | Automated Multi-Dataset Analysis |
| au | Astronomical Unit (149,597,870.7 km) |
| BOL | Beginning of Life |
| CEST | Central European Standard Time |
| CNES | Centre National d'Etudes Spatiales |
| COSPAR | Committee on Space Research |
| DHAWG | Data Archiving Working Group |
| DSN | Deep Space Network |
| ECA | Evolution Cryotechnique type A |
| ESA | European Space Agency |
| ESAC | European Space Astronomy Center |
| ESOC | European Space Operation Center |
| ESTEC | European Space Technology Center |
| ESTRACK | European Space Tracking network |
| GALA | Ganymede Laser Altimeter |
| GAM | Gravity Assist Manoeuvre |
| HAA | High Accuracy Accelerometer |
| HGA | High Gain Antenna |
| JANUS | Jovis, Amorum ac Natorum Undique Scrutator |
| JEO | Jupiter Europa Orbiter |
| JGO | Jupiter Ganymede Orbiter |
| J-MAG | JUICE Magnetometer |
| JMC | JUICE Monitoring Camera |
| JUICE | JUpiter Icy moons Explorer |
| MAJIS | Moons And Jupiter Imaging Spectrometer |
| MGA | Medium Gain Antenna |
| MOC | Mission Operations Centre |
| NAVCAM | Navigation Camera |
| OST | Outer Space Treaty |
| PEP | Particle Environment Package |
| PI | Principal Investigator |
| PRIDE | Planetary Radio Interferometry and Doppler Experiment |
| RADEM | RADiation Environment Monitor |
| RIME | Radar for Icy Moons Exploration |
| RPWI | Radio and Plasma Wave Investigation |
| SOC | Science Operation Centre |
| SOWG | Science Operation Working Group |
| SPC | Science Programme Committee |
| SWI | Sub-mm Wave Instrument |
| SWT | Science Working Team |
| UVS | UV Spectrograph |
| UTC | Universal Time Coordinated |
| | The full list of JUICE acronyms can be found at: https://www.cosmos.esa.int/web/JUICE/JUICE-acronyms |

## 1 Introduction

JUICE (JUpiter ICy moons Explorer) is the first European mission to Jupiter and its icy Moons (Grasset et al. 2013a). The spacecraft was launched on 14 April 2023 from the European Spaceport in French Guiana by the Ariane 5 launcher. The first concepts for a new mission to Jupiter and its moons date back to 2007, at around the same time as the Juno mission (focussed on Jupiter itself) was in development in the USA. It was inspired by the spectacular observations of the Galilean moons returned by NASA's two Voyager flybys in 1979 (see Kohlhase and Penzo 1977; Stone and Lane 1979; Smith et al. 1979; Evans 2022) and by NASA's Galileo mission (Johnson et al. 1992) that orbited Jupiter from December 1995 to September 2003. At that time, ESA was gaining considerable experience in giant planet exploration thanks, among other, to the very successful joint mission to Saturn and its moon Titan, the NASA/ESA/ASI mission called Cassini-Huygens for the orbiter and the probe names, respectively (Matson et al. 2002; Coustenis and Hirtzig 2009; Lebreton et al. 2009; Dougherty et al. 2009). The international mission orbited Saturn from July 2004 to September 2017, executed 127 flybys of Titan, and revolutionised our understanding of the Saturn system (Spilker 2019). The Huygens probe descended on Titan in January 2005 and delivered a unique set of data of the atmosphere and surface of the satellite during several hours (Lebreton et al. 2005).

Building on this experience, conditions were favourable for a new joint ESA-NASA mission targeting Jupiter's system. A proposal for such a mission, named Laplace was first submitted to ESA in 2007 (Blanc et al. 2009). This was followed by a joint study during 2008–2010, by the two agencies that recommended a mission with two spacecraft (orbiters), one with Ganymede as the main target led by ESA (Jupiter Ganymede Orbiter, JGO), and one targeting Europa led by NASA (Jupiter Europa Orbiter, JEO). The mission was called the Europa Jupiter System Mission (EJSM, formerly EJSM-Laplace, Clark et al. 2009). Unfortunately, NASA soon realised that they could not afford the mission in the considered timeframe and, in addition, an orbiter around Europa was deemed unfeasible due to the challenging effects of the Jovian radiation field on the proposed instruments. Following that decision in 2010, a proposal was made to ESA to move forward with a European orbiter only (the JGO), enhanced to cover not only Ganymede but also the other two icy moons, Europa and Callisto as well (Yellow Book, ESA/SRE(2011)18). It was subsequently renamed as JUICE for JUpiter ICy moons Explorer. After a period of consolidation during the preliminary study phase of the overall mission objectives and scientific requirements, the mission was eventually selected within the Cosmic Vision 2015–2025 plan (Bignami et al. 2005) by the ESA Science Programme Committee (SPC) in 2012 for the detailed study involving, in a competitive manner, the two leading European space industries: Thales Alenia Space and Airbus Défense and Space. The deep study lasted two years and led to the selection of the instruments to be flown, the establishment of multilateral agreements with the European Member States and NASA providing the instruments, a consolidated set of technical requirements and a mature preliminary design of the spacecraft to define the overall schedule and cost at completion. On that basis, the adoption of the program, namely the final decision to proceed with the implementation, was made by the ESA SPC in November 2014, based on the JUICE Red Book submitted in September 2014 (ESA/SRE(2014)1). The prime industrial contractor, Airbus Défense and Space, was selected a few months later. The overall timeline of the programme until the launch is shown in Table 1.

This article provides a post-launch assessment of the mission, summarising its scientific objectives (Sect. 2), planetary protection drivers (Sect. 3), trajectory design (Sect. 4), spacecraft architecture and launcher (Sect. 5), instrumentation suite (Sect. 5), ground-segment (Sect. 5), inter-mission collaborations (Sect. 6), and synergies with Earth-based observatories (Sect. 7).

## 2 Scientific Objectives of the Mission

JUICE pursues a set of highly ambitious goals (e.g., Grasset et al. 2013a). These include investigating Jupiter, its moons and rings, to cast new light on the formation, evolution and stability of planetary systems, and to reveal the structure and processes shaping the atmosphere and magnetosphere of Jupiter. Indeed, Jupiter's magnetosphere is the largest structure in our solar system and a major component in the processes at play in the Jovian system. Moreover, JUICE will also study the conditions that may have led to the emergence of habitable environments among the Jovian icy satellites. Habitable environments are defined as planetary bodies that could support, or might have supported, organic life and hosting liquid water, essential elements and energy sources.

**Table 1** Key milestones of the JUICE programme up to launch

| Milestone | Date |
|---|---|
| ESA call for proposals for an L-Class mission within the Cosmic Vision plan | 3/2007 |
| JUICE assessment study report (Yellow Book) | 12/2011 |
| Mission selection | 5/2012 |
| Payload selection | 7/2/2013 |
| JUICE definition study report (Red Book) | 9/2014 |
| Mission adoption | 6/11/2014 |
| Industrial activities kick-off | 28/7/2015 |
| Spacecraft System Requirement Review (SRR) | 1/3/2016 |
| Instrument's Preliminary Design Reviews (PDRs) | 8/4/2016 – 25/9/2017 |
| Spacecraft Preliminary Design Review (PDR) | 1/3/2017 |
| Instrument's Critical Design Reviews (CDRs) | 18/10/2017 – 2/7/2019 |
| Spacecraft Critical Design Review (CDR) | 4/3/2019 |
| Spacecraft integration and testing | 7/2020 – 12/2022 |
| Spacecraft Qualification and Flight Acceptance Review (QR/FAR) | 18/1/2023 |
| Spacecraft shipment to Kourou | 9/2/2023 |
| Launch | 14/4/2023 |

To this end, the mission will focus on three icy Galilean moons: Ganymede, Europa, and Callisto. The reason for this choice is that they are thought to harbour vast internal oceans: each of the icy moons may contain more water than all the oceans on Earth (e.g., Lunine 2017; Hand et al. 2020). Identifying and further characterising liquid water, its location and volume, is crucial in the search for habitable worlds beyond Earth and in the search for life as we know it. JUICE will especially focus on Ganymede, the largest moon in our solar system. With its supposed saltwater ocean buried under a shell of mostly ice (Kivelson et al. 1996; Paty et al. 2008), Ganymede provides a natural laboratory to study the potential habitability of icy moons and water worlds (Saur et al. 2015). Ganymede is also the only moon known to have its own magnetic field (Kivelson et al. 1996). In general, having an internal magnetic field is presently a rare feature for solid bodies in our solar system (Earth and Mercury are the only other terrestrial bodies with an internal magnetic field in our solar system); and Ganymede's magnetic field plays a fundamental role in protecting the moon's surface from radiation. It is important to point out that JUICE itself will not look for direct signs of life, but will rather focus on understanding whether the environmental conditions of the moons are suitable for the emergence of life, a prerequisite to future missions aiming at searching for life in the oceans of moons that will have been characterised as habitable (e.g., Hendrix et al. 2019; Hand et al. 2020).

The scientific objectives of JUICE are described in Grasset et al. (2013a), Coustenis et al. (2021), Grasset et al. (2023), in the Definition Study Report (the 'Red Book'; ESA 2014b), and in the JUICE Science Requirements Document (ESA 2014a). Companion articles (Van Hoolst et al. 2024; Tosi et al. 2024a; Masters et al. 2025; Fletcher et al. 2023 and Denk et al. 2026) describe the science objectives in more detail. The mission is built around two overarching science themes: "the emergence of habitable worlds around gas giants", also expressed as "exploring the habitable zone: Ganymede, Europa, and Callisto", and "the Jupiter system as an archetype for gaseous giant planets". Each theme is associated with three observational targets: Ganymede, Europa, and Callisto for the first; Jupiter's atmosphere, magnetosphere and its satellite and ring systems for the second. These six targets are allocated an agreed share of the mission's data volume (Altobelli et al. 2026, this collection). Table 2 presents the nineteen scientific objectives covering the two themes and their associated targets, together with a selection of the scientific questions that JUICE will address (Assessment Study Report, 'Yellow Book'; ESA 2011).

JUICE will perform comprehensive system-level science at Jupiter. The mission places particular emphasis on the three icy moons and their potential habitability. These three of the four Galilean moons span a wide range of internal structures, from predominantly rocky bodies to those rich in ice, and fall into two broad classes: the mainly rocky moon Europa, and the mixed rock-ice moons Ganymede and Callisto (Nimmo and Pappalardo 2016). Europa consists largely of rock overlayed by a relatively thin ice shell and a subsurface ocean (e.g., Kivelson et al. 2000). Ganymede is thought to be fully differentiated, with a rocky mantle and metallic core beneath a thick ice shell and subsurface ocean (Saur et al. 2015); Callisto is likely only partially differentiated, with a deep interior composed of mixed ice and rock, and possibly also a subsurface ocean beneath a thick ice shell (Khurana et al. 1998). To put the icy moons within their broader system context, and to better understand the Galilean satellites as interconnected components of the Jovian system, the observation strategy will include:

- a comparative study of Ganymede, Callisto, and Europa, with particular focus on characterising Ganymede as a planetary body and potential habitat;
- a full spatio-temporal characterisation of Jupiter's rapidly rotating magnetosphere, as well as the meteorology, chemistry, and vertical structure of its atmosphere, from the cloud tops upward, which drive the magnetosphere into its diurnal rotation via frictional coupling in the thermosphere-ionosphere;
- an investigation of coupling processes throughout the Jupiter system, with emphasis on tidal interaction, electrodynamic coupling and their mutual interplay.

## 2.1 Ganymede

Here, we provide a brief overview of the science objectives at Ganymede, spanning from its deep interior to its surrounding space environment (Van Hoolst et al.; Tosi et al.; Masters et al., this collection). A central goal of JUICE is to investigate Ganymede's icy shell, subsurface ocean, and deep interior. The thickness of the ice shell will be constrained through measurements of the time variability of the (degree 2) gravity field, the amplitudes of periodic tidal surface deformation, and the electromagnetic induction response generated within the electrically conductive subsurface ocean. These datasets will not only reveal the average thicknesses of the outer ice shell and the ocean but will also constrain the ocean's salt content. Additional insights will come from measurements of the gravity field up to degree 12, the surface topography, physical librations and the orientation of the spin poles. Together, these results will provide information on lateral variations in the shell's thickness and density, on the average thickness of the high-pressure ice mantle, and on ocean flow, heat transfer through the hydrosphere, and key factors governing subsurface habitability. The mission will also refine our understanding of Ganymede's outer shell and interior structure by determining the moon's centre of mass position relative to Jupiter. This links the orbital mechanics of the Laplace resonance to Ganymede's internal structure. A comprehensive characterisation of Ganymede's intrinsic magnetic field will additionally provide essential constraints on the conditions needed to generate and sustain a dynamo. JUICE will also investigate the structure of the icy shell, including the potential presence of shallow liquid water reservoirs; analyse the crust's composition, structure, and physical properties;

**Table 2** Synthesis of JUICE science goals: two main themes (in bold), six targets (left column) and nineteen scientific objectives (middlecolumn). Selected scientific questions to be addressed by JUICE are listed in the right-hand column

| Explore the habitable zone: Ganymede, Europa, and Callisto | | |
|---|---|---|
| Ganymede as a planetary object and possible habitat | • Characterise the extent of the ocean and its relation to the deeper interior<br>• Characterise the ice shell<br>• Determine global composition, distribution, and evolution of surface materials<br>• Understand the formation of surface features and search for past and present activity<br>• Characterise the local environment and its interaction with the Jovian magnetosphere | • Is there liquid water on Ganymede? If so, what is its spatial distribution? How thick is the ice layer and what are the properties of the liquid?<br>• What is the role of tidal heating in the evolution of Ganymede?<br>• Is there material exchange between the deep interior, ocean, ice shell, and surface, and how is this related to non-water-ice components at the surface?<br>• How is the geological evolution related to the impact, tectonic and cryovolcanic history and how is it correlated with differentiation processes and stages?<br>• What are the characteristics of Ganymede's various magnetic field components (dynamo, induced) and mini-magnetosphere?<br>• How does the local plasma environment affect the surface, and exosphere/ atmosphere? |
| Europa's recently active zones | • Determine the composition of the non-ice material, especially as related to habitability<br>• Search for liquid water under the most active sites<br>• Study the recently active processes | • Is there liquid water in the shallow icy crust of Europa?<br>• What is the minimal thickness of the icy crust over the most active regions?<br>• What is the composition and origin of the non-ice material, and how is it related to recent geological activity, shallow subsurface water, and habitability? |
| Callisto as a remnant of the early Jovian system | • Characterise the outer shells, including the ocean<br>• Determine the composition of the non-ice material<br>• Study the past activity | • Is there any liquid water embedded in Callisto's icy crust?<br>• Is Callisto in hydrostatic equilibrium? What is the state of differentiation? How is the incomplete separation of ice and rock related to the formation of a subsurface ocean?<br>• What does the distribution of craters tell us about the evolution of the impactor population through time?<br>• What is the origin and composition of Callisto's dark non-water-ice material and volatiles, and what do they reveal about surface processing and the evolution of the outer ice shell? |

**Table 2** (*Continued*)

| | | |
|---|---|---|
| Explore the Jupiter system as an archetype for gas giants | | |
| The Jovian atmosphere | • Characterise the atmospheric dynamics and circulation<br>• Characterise the atmospheric composition and chemistry<br>• Characterise the atmospheric vertical structure | • What dynamical processes are involved in the redistribution of deposited solar energy, internal potential energy and material tracers within the Jovian atmosphere and between the different layers?<br>• How is the spatial variation of condensable species and disequilibrium gases related to the meteorology of Jupiter's weather layer?<br>• What processes govern the production of haze within the polar vortices, and what drives the hemispheric asymmetry? |
| The Jovian magnetosphere | • Characterise the magnetosphere as a fast magnetic rotator<br>• Characterise the magnetosphere as a giant accelerator<br>• Understand the moons as sources and sinks of magnetospheric plasma | • How do the global magnetospheric structure and activity depend on solar wind effects and mass-loading processes?<br>• Where do the high energy particles in the Jovian radiation belts come from? How are they produced in the most intense radiation environment in the Solar System? How do they affect moons (their surfaces, tenuous atmospheres/exospheres) and what are the effects in terms of habitability? |
| The Jovian satellite and ring systems | • Study Io's activity and surface composition<br>• Study the main characteristics of rings and small satellites | • How will the surface of Io, as well as the neutral cloud and plasma torus, change over the course of the Juice mission?<br>• What is the 3D structure of Jupiter's ring system?<br>• How similar are the small inner moons to each other, what are they made of, and how have they been formed?<br>• Are there statistically relevant, non-random patterns in the rotation periods and pole-axis orientations of the Irregular satellites? |

study interactions between the surface and subsurface ocean; and assess correlations between surface features (geological and compositional) and subsurface structure.

The mission will deliver a major advance in Ganymede's geology and chemical composition. Ganymede's surface shows two terrain types—dark terrain and light terrain possibly remodelled by tectonism, therefore implying past or ongoing internal activity, probably involving water (Moore et al. 2004). Through global mapping in the UV, visible, infrared and sub-millimetre, including high resolution imaging of key terrains and the production of digital terrain models at various scales, JUICE data will place constraints on these theories. One of the objectives is to determine how internal and external processes (impact, tectonics, and cryovolcanism) have shaped the moon's surface over time. JUICE will improve surface age estimates through crater population analyses and will study surface chemistry and its evolution, evaluating contributions from endogenic activity, external processes, and exchange with the subsurface, as well as the sources of Ganymede's tenuous atmosphere. A list of regions of interest is documented in Stephan et al. (2021).

Since the subsurface ocean will be detected through magnetic induction, a crucial aspect of the mission is to characterise Ganymede's space environment and its electrodynamic coupling with Jupiter, to separate internal and external contributions to Ganymede in magnetic field variations. JUICE will examine the interplay between the intrinsic magnetic field, the fields induced within the ocean, and the Jovian magnetosphere (Saur et al. 2015). The mission will conduct both remote sensing and in situ observations of all of Ganymede's external plasma envelopes (the ionosphere, exosphere, and magnetosphere) including detailed studies of its auroras. JUICE will also perform observations of Ganymede's atmosphere resulting from the interaction of the particle and photon environment with its surface (Galli et al. 2025).

## 2.2 Europa

During its two close flybys of Europa, JUICE will concentrate on characterising the moon's non-icy materials, searching for liquid water beneath potentially active regions, and investigating active processes (Schmidt et al. 2011). During these flybys, JUICE will also measure the composition of Europa's atmosphere.

A key objective is to map the distribution and composition of Europa's non-ice material and to relate it to the moon's interior structure, subsurface ocean, and the effects of space weathering. JUICE will also search for subsurface liquid reservoirs (Vorburger and Wurz 2021) and probe the interface between the ice shell and the underlying ocean. These investigations will be complemented by topographic measurements, gravity field determination, and constraints on Europa's global shape. The mission will further explore active processes, including the possible detection of plumes, the characterisation of the local space environment (ionosphere and exosphere), the study of the Europa torus, and the identification of surface changes through comparisons with previous datasets.

## 2.3 Callisto

Callisto is an intriguing moon: its heavily cratered surface indicates no major geological activity for billions of years, yet Galileo measurements revealed an induced magnetic field possibly consistent with the presence of a subsurface ocean (Zimmer et al. 2000; Cochrane et al. 2025).

During its flybys, JUICE will investigate Callisto's icy crust and underlying liquid layer and will characterise its space environment to precisely determine the moon's magnetic induction response. It will also refine measurements of Callisto's low-degree gravity field and moments of inertia, thereby providing strong constraints on its deep interior structure and degree of differentiation.

Because the inventory of non-water-ice constituents on Callisto remains to be fully established, JUICE will characterise the surface composition in detail and relate it to geological processes and the long-term evolution of the moon. Together with the imaging of impact craters JUICE will unveil the history of Jovian and Solar system formation.

Callisto's atmosphere of primarily $O_2$, $CO_2$, $H_2O$, and $H_2$ is thought to be denser than Ganymede or Europa, owing to its surface composition, atmospheric production processes, and distinct magnetospheric environment within the Jupiter's system.. The regime is collisional as opposed to a surface-bounded exosphere (Carberry Mogan et al. 2020, 2021, 2022), and results in a relatively high-conductivity ionosphere that efficiently diverts much of the incoming plasma flow from the magnetosphere. JUICE's remote sensing and in situ instruments will map the structure of Callisto's atmosphere, providing an interesting contrast with the plasma-atmosphere investigations at Ganymede and Europa.

## 2.4 Jovian Atmosphere

Our understanding of Jupiter's atmosphere is still incomplete, with many fundamental questions to be resolved. Studying the atmospheric composition and distribution of material with depth provides crucial insights into the formation and evolution of giant planets, and more broadly, into the nature of exoplanetary systems. JUICE will investigate the three-dimensional spatial and temporal variability of Jovian atmosphere dynamics, chemistry, and structure (full details can be found in Fletcher et al. 2023).

A key objective is to observe time-variable phenomena across a wide range of latitudes and local solar times, over a long temporal baseline and broad spectral range. Near-simultaneous multispectral observations will link environmental parameters, such as temperatures, chemical composition, and winds, to changes in visible albedo and meteorological activity. With repeated mapping from equator to pole, JUICE will identify the dynamical processes that drive Jupiter's atmospheric variability. The mission will study atmospheric circulation and meteorology from the cloud-forming troposphere up to the thermosphere, relating these layers to the planet's deep interior and to the influence of its coupling to the magnetospheric environment.

JUICE will probe the three-dimensional distribution and variability of key stratospheric species to investigate middle-atmosphere circulation and photochemistry. In particular, submillimetre spectroscopy will reveal motions and circulation patterns within the middle atmosphere for the first time. JUICE will study exogenic constituents (e.g., $H_2O$) deposited in the stratosphere to identify external material sources; trace disequilibrium species (e.g., $PH_3$) associated with localised atmospheric features; and endogenic volatiles (e.g., ammonia and water) to assess the role of moist convection in cloud formation, lightning, and chemical processes.

The mission will also characterise vertical atmospheric structure and the coupling mechanisms that transport energy and momentum, such as wave propagation, ion drag, and meridional transport, providing complete latitudinal and local-time coverage. Jupiter's banded structure, vortices, and storms are known to vary over a range of timescales, from days to years. The long temporal baseline of the JUICE mission will enable the study of how these processes evolve and respond to internal forcing and external drivers over long periods of time.

A major theme of the JUICE science case is that Jupiter's interior, atmosphere, and magnetosphere form a tightly connected system that cannot be understood in isolation. JUICE, complemented by the legacy of the Juno mission, will therefore investigate the links and feedback loops between all components of this coupled environment (Fletcher et al. 2023).

## 2.5 Jovian Magnetosphere

JUICE will investigate Jupiter's vast and complex magnetosphere by characterising its principal regions: the inner magnetosphere dominated by the planetary magnetic field, the middle magnetosphere shaped by the magnetodisc and its control of large-scale fields, flows and plasma populations, and the outer magnetosphere where solar wind forcing and system-level variability are strongest. Along its trajectory, JUICE will examine the structure and dynamics of the magnetodisc, key plasma processes, sources and sinks of particles, their composition and transport, and the variability of these phenomena.

A major objective is to characterise the electrodynamic coupling within the Jovian system, linking the magnetosphere, ionosphere, and thermosphere to the Galilean moons and their tori, smaller satellites, and the solar wind. This objective will be addressed via detailed observations of auroral emissions, plasma flows, magnetic field variations, satellite footprints, particle precipitation onto moon surfaces and their effects on surface properties, the contribution to the atmosphere, and in particular, interaction of the Jovian magnetic field with that of Ganymede. JUICE will also investigate particle acceleration mechanisms and synchrotron emissions.

While JUICE's contribution to magnetospheric science will be broad, two areas stand out (Masters et al. 2025). The mission will provide unprecedented insight into Jupiter's equatorial magnetosphere, including the region of corotation breakdown. And it will deliver transformative observations of Ganymede's magnetosphere, unique in the Solar System as the only moon with an intrinsic magnetic field, thereby offering a unique opportunity to study a magnetosphere embedded inside a magnetosphere.

## 2.6 Other Satellites and Ring Systems

Besides the three icy Galilean moons, the Jovian satellite system includes the active moon Io, the four small inner moons Metis, Adrastea, Amalthea, and Thebe, and the numerous small irregular (outer) moons, as well as a faint ring system. JUICE will observe Io and its neutral cloud and plasma torus from distance, allowing a monitoring of its volcanic activity and surface changes, to characterize its atmospheric composition and nightside densities as well as to improve its orbit determination. The inner moons will be imaged to infer their physical shape, colours, large surface structures, and bulk composition, while astrometric measurements will be performed to improve the ephemerides. The 3-D structure, physical and chemical properties of Jupiter's rings and its four main components will be characterised remotely and with radio-occultations, as well as the ring-small satellites connections. Although they will mostly appear unresolved, the outer irregular moons of Jupiter will be observed to determine basic physical properties like rotation periods, pole directions, object dimensions and shapes, plus for improvement of orbit shapes. A close flyby of the small irregular moon Kallichore (J XLIV) has been identified as an opportunity and is currently under discussion (Boutonnet and Rocchi 2026; Denk et al. 2026, this collection). Discovery of new satellites might be possible with JUICE. Io and the minor moons are discussed in depth by Denk et al. (2026).

### 2.7 Cruise Phase Science

Although cruise-phase science is not part of the mission baseline, scientific observations are nonetheless conducted when permitted by operational constraints and available resources, as enabled by the mission trajectory.

Calibrations and tests are performed during the Earth and Moon gravity assists. These measurements and observations have already yielded valuable scientific insights (e.g., Poulet et al. 2026; Agostini et al. 2026; Tubiana et al. 2026; Hartogh et al. 2027; Langevin et al. 2026; Guerlet et al. 2026; Hueso et al. 2026; Oliva et al. 2026; Cavalié et al. 2026; D'Aversa et al. 2026; Jarchow et al. 2026; Zambon et al. 2026; Tosi et al. 2026; Molera Calvés et al. 2026; Lucchetti et al. 2026; Brandt et al. 2026; Pinto et al. 2026a; De Filippis et al. 2026; Moreno et al. 2026; Molyneux et al. 2026; LaMoury et al. 2026; Fränz et al. 2026; Dialynas et al. 2026; Nishiyama et al. 2026; Zambon et al. 2026; Stenberg Wieser et al. 2026; Vallat et al. 2026; Seignovert et al. 2026).

The cruise phase offers also an opportunity to conduct measurements valuable to the heliophysics community (e.g. Witasse et al. 2017; Knutsen et al. 2021; Hadid et al. 2021; Strauss et al. 2024; Sánchez-Cano et al. 2025; Nakamura et al. 2025; Kinoshita et al. 2026). It enables the collection of additional data on solar wind structures and space-weather events, Jovian electrons, galactic cosmic rays' modulation, solar radio-bursts, and contributes to multi-point observations that are essential for advancing heliophysics research. The JUICE trajectory is particularly favourable for such investigations, especially between 1.4 astronomical units (beyond which in-situ measurements become easier to schedule due to thermal constraints by the spacecraft are more relaxed) and 5 au. Remote imaging of the heliospheric interaction with the interstellar medium and observations of Hydrogen and Helium skyglow will be planned. The radiation monitor is near-continuously operating and provides a useful data set to the community in term of electron, proton, and heavy ion fluxes. Observations and analysis of phase scintillation of spacecraft signal can allow useful information on the properties of the interplanetary plasma to be derived (Molera Calvés et al. 2014; Said et al. 2025; Edwards et al. 2025).

Extending the in-situ cruise observations by 15 months prior to Jupiter Orbit Insertion (JOI) will enable coordinated investigations with Europa Clipper. As mentioned in Sect. 7, this two-spacecraft configuration (combining JUICE's approach to Jupiter with Europa Clipper located inside the magnetosphere) will allow joint studies of solar wind-magnetosphere interactions, auroral processes, and the impact of solar wind variability on the local plasma environments of the moons.

Cruise phase can also be used to study asteroids and comets, when opportunities materialise. In the case of JUICE, opportunities to flyby an asteroid have been studied (Avdellidou et al. 2021; Agostini et al. 2022), but a close encounter was ultimately not pursued to preserve sufficient propellant margins for the nominal science phase.

The interstellar object 3I/ATLAS passed through our solar system in 2025 (Seligman et al. 2025; Eubanks et al. 2025), and due to its exceptional nature, a dedicated observation campaign was conducted in November 2025 immediately after perihelion, yielding a unique dataset. The collected data are currently under analysis (e.g., Bockelée-Morvan et al. 2026). Europa Clipper also observed 3I/ATLAS at the same time.

The in-situ detection of cometary dust trails and the analysis of meteoroid-stream particles represent another valuable scientific opportunity during the cruise phase. Because these particles can be traced back to their parent bodies, they offer a novel means of remotely assessing the composition of comets and asteroids (Krüger et al. 2024). Several dust-trail crossings are predicted along JUICE's trajectory, and dedicated attempts to detect them will be carried out. Interplanetary and interstellar dust detection (Sterken et al. 2019, 2023) will be attempted whenever possible, to produce a dust count temporal profile. Opportunities for crossing plasma tails of comet will also be checked (e.g. Jones et al. 2000).

The cruise phase offers the possibility to carry out classical tests of general relativity by measuring the relativistic time delay and the frequency shift of radio signals during superior solar conjunctions (di Stefano et al. 2022). These experiments, under preparation, aim to provide the most precise determination of the post-Newtonian parameter. Because JUICE will perform these measurements at large heliocentric distances, where solar-wind disturbances are expected to be lower, the resulting data are anticipated to improve upon the experiments conducted with Cassini (Bertotti et al. 2003) and BepiColombo (Iess et al. 2021).

The JUICE cruise-phase navigation data will support improvements in planetary ephemerides and can be used to provide new constraints on planetary bodies (e.g. Fienga et al. 2020; Di Ruscio et al. 2020).

## 3 Planetary Protection

Spacecraft venturing into the Solar System must ensure that they do not introduce terrestrial biological contamination to planets and moons that may harbour present life or traces thereof. The Committee on Space Research (COSPAR 2026; Olsson-Francis et al. 2026) has established a Planetary Protection Policy to guide compliance with the United Nations Outer Space Treaty (OST). Adherence to these planetary protection requirements constitutes a non-legally binding obligation but is currently considered as a responsibility by the 118 nations having ratified the OST, in particular articles IX and VI. The European Space Agency has adopted this policy and implemented the necessary measures to ensure that its missions comply. For the JUICE mission, planetary protection requirements impose constraints on the probability of unintentional impacts with Mars and Europa (Tobie et al. 2026). JUICE will conduct two flybys of Europa, which is classified by COSPAR's Planetary Protection Panel as a body with the potential to harbour life (Boutonnet and Martens 2020). Consequently, this mission phase must comply with Planetary Protection Category III requirements, ensuring that the probability of accidental contamination of the subsurface ocean remains below $10^{-4}$. To avoid lengthy analyses, it was decided to put all the burden solely on the impact probability with Europa. The overall impact probability is allocated between short-term and long-term risks. Short-term risk refers, for example, to a contingency scenario such as entry into safe mode followed by an unintended collision (Boutonnet and Martens 2020; Tobie et al. 2026). Long-term risk encompasses scenarios such as complete spacecraft failure leading to an uncontrolled trajectory evolution and eventual impact. The long-term impact probability has been computed for JUICE through statistical analyses propagated over a 1000-year timeframe. The short-term risk has been computed for the period of the tour just before both Europa flybys. It was demonstrated that the sum of both probabilities is lower than $10^{-4}$. Regarding impacting Ganymede, it was demonstrated that the probability of contamination of subsurface liquid reservoirs is well below the required threshold (Grasset et al. 2013b; Tobie et al. 2026). The final impact of JUICE on Ganymede at the end of mission is therefore compliant with planetary protection requirements. For Mars, there is no Mars gravity assist during the cruise; it was nevertheless demonstrated by analysis that the COSPAR requirement is met.

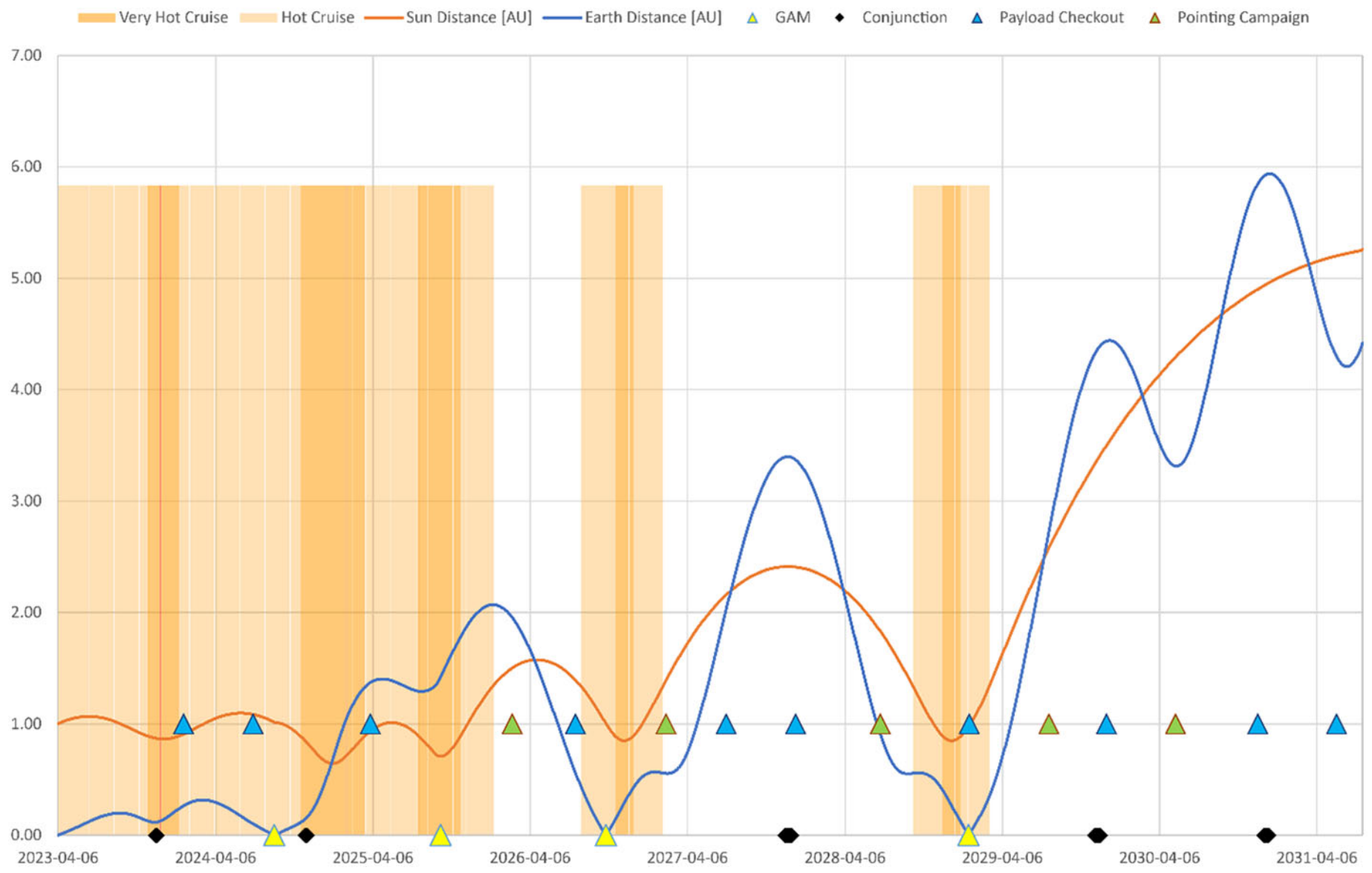

**Fig. 1** Cruise phase overview: time (X axis), distances to Earth (blue) and to the Sun (red/orange) in astronomical units (Y axis), gravity assists manoeuvres (GAM) (yellow triangles; the Venus flyby is the second triangle from the left), Solar superior conjunction (black diamonds), payload checkouts (blue/green triangles). The hot and very hot phases are displayed with vertical orange bars

## 4 The JUICE Trajectory

The JUICE trajectory is divided into three main phases: the eight-year cruise from Earth to Jupiter, the three-year Jupiter tour, and the one-year orbital phase around Ganymede. The trajectory design, resulting from more than 20 years of mission analysis, had to satisfy a wide range of requirements, including scientific objectives (multiple targets), illumination constraints, propellant and navigation limitations, launch opportunities, eclipse duration limits, radiation environment considerations, planetary protection requirements, operational constraints, and the expected mission lifetime. The trajectory design is documented in full detail in Boutonnet et al. (2024). The Jupiter tour and Ganymede orbital phase are expected to evolve in the coming years, and any resulting changes will be documented. The Ariane 5 ECA launcher injected the spacecraft directly onto an escape trajectory. The launcher's performance, combined with an optimal launch date near the middle of the launch window, provided additional propellant margin. This margin, together with the spacecraft dry-mass margin at launch that allowed for fully filling the propellant tanks, made it possible to reinstate a final 200 km-altitude orbital phase around Ganymede in the post launch baseline trajectory that was previously limited to the 500 km-altitude circular orbits. This final orbit had previously been removed during the study phase due to mass-margin constraints. The cruise phase to Jupiter, illustrated in Figs. 1 and 2, lasts eight years and includes four gravity-assist flybys, as listed in Table 3. At the time of writing, JUICE has completed the first two flybys and is currently *en route* to its second Earth gravity assist. During the cruise phase, periodic payload checkouts are performed to verify instrument health, monitor performance, and conduct necessary calibration activities. Critical calibrations, operating tests, and scientific observations were carried out during the Earth–Moon gravity-assist flybys.

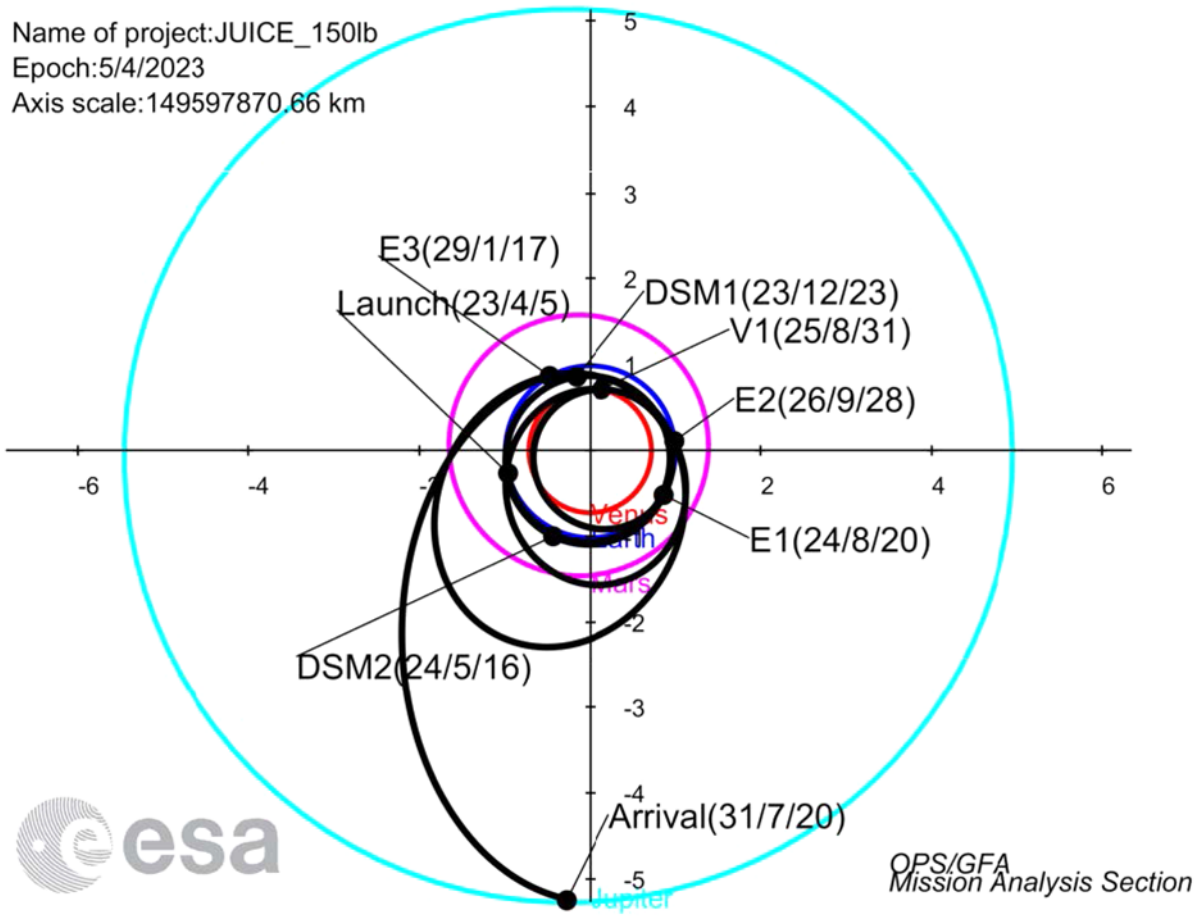

**Fig. 2** Cruise trajectory from launch to Jupiter orbit insertion (from Boutonnet et al. 2024). DSM: Deep Space Manoeuvre. E1: Moon-Earth flybys; E2, E3: Earth flybys. V1: Venus's flyby. The Sun is at the centre. Mars' orbit is shown in pink, Earth's orbit in blue, Venus' orbit in red

**Table 3** List of the four gravity assist flybys during the JUICE cruise phase, including the altitudes at closest approach

| Gravity assist | Date | Altitude at closest approach (km) |
|---|---|---|
| Lunar-Earth | 19–20 August 2024 | Moon: 752, Earth: 6839 |
| Venus | 31 August 2025 | 5098 |
| Earth | 28 September 2026 | 8639 |
| Earth | 17 January 2029 | 4640 |

No payload operations were planned during the Venus flyby due to strict operational and thermal constraints; only the radiation monitor was activated.

Tables 4 and 5 provide information on the timeline and main characteristics of the trajectory. Figure 3 displays the trajectory in the Jupiter system.

In July 2031, the spacecraft will be inserted into orbit around Jupiter via a capture manoeuvre. An initial Ganymede swing-by, performed at an altitude of 400 km approximately 13 hours before Jupiter orbit insertion, will reduce the required engine burn duration. A flyby of the moon Kallichore during the first highly elliptical orbit, about two and a half months after Jupiter capture, is currently under evaluation and could be incorporated into an updated baseline trajectory. Following a sequence of Ganymede swing-bys that progressively reduce the orbital energy, inclination, and hyperbolic excess velocity, the spacecraft will enter a one-month Europa science phase, including two flybys at a minimum altitude of 400 km. The number of Europa flybys is intentionally limited to minimise radiation exposure. This phase will be followed by a Jupiter high-latitude segment, during which Callisto gravity assists are used to raise the orbit inclination to 35 degrees. Subsequent Callisto flybys will then return the spacecraft to the equatorial plane. A further series of Ganymede and Callisto flybys will prepare the spacecraft for gravitational capture by Ganymede, planned for the end of October 2034. The capture manoeuvre will nevertheless be applied to lower the apocentre to approximately 10,000 km.

The orbital phase around Ganymede consists of an initial elliptical sub-phase, with an orbital period of roughly 12 hours, followed by two low-altitude circular sub-phases at approximately 500 km and 200 km altitude as mentioned. The elliptical phase will last about

**Table 4** Top-level timeline of the JUICE trajectory, from launch to end of mission, for the current baseline

| Events | Date |
|---|---|
| Launch | 14 April 2023 |
| 1st Ganymede flyby | 21 July 2031 |
| Jupiter orbit insertion | 21 July 2031 |
| Energy reduction phase | 21 July 2031 – 20 June 2032 |
| Putative Kallichore flyby | 5 October 2031 |
| 2nd Ganymede flyby | 6 February 2032 |
| 1st Callisto flyby | 21 June 2032 |
| Europa science phase | 21 June 2032 – 13 August 2032 |
| Two Europa flybys | 2 and 16 July 2032 |
| Jupiter inclined phase | 14 August 2032 – 1 November 2033 |
| Transfer to Ganymede phase | 2 November 2033 – 30 October 2034 |
| Last Callisto flyby | 24 June 2034 |
| Last Ganymede flyby | 29 September 2034 |
| Ganymede orbit insertion | 31 October 2034 |
| End of nominal mission | 12 October 2035 |

**Table 5** Selection of characteristics of the JUICE nominal science phase

| | |
|---|---|
| Duration of the Jupiter tour | 3 years and 3 months |
| Duration of the Ganymede orbital phase | 11.5 months |
| Number of perijoves | 62 |
| Number of Europa flybys | 2 |
| Number of Ganymede flybys | 11 |
| Number of Callisto flybys | 23 |
| Maximum inclination of the orbit during the tour | 35 degrees |

150 days. During this period, the eccentricity first decreases rapidly, remains close to zero for some time at an altitude of about 5000 km, and then increases again under Jupiter's gravitational perturbations. The orbit will be phased such that communication passes occur on the nightside, enabling imaging operations on the dayside. The transition from the elliptical phase to the circular phases will be achieved through three manoeuvres spaced four days apart. The first circular orbit at around 500 km altitude will last approximately 160 days, after which the spacecraft will be transferred close to a 200 km orbit for roughly one month. A mission extension may be possible, depending on scientific priorities, spacecraft and instrument performance, remaining propellant, available funding, and ESA decision processes. In the current baseline scenario, the spacecraft will ultimately impact Ganymede's surface in an uncontrolled manner. Should a controlled impact be required, an intentional de-orbit manoeuvre would need to be performed.

## 5 JUICE Mission Elements

The JUICE programme is composed of several elements, some of which are specific for the purpose of the mission, which had to be developed in a timely manner. This is shown

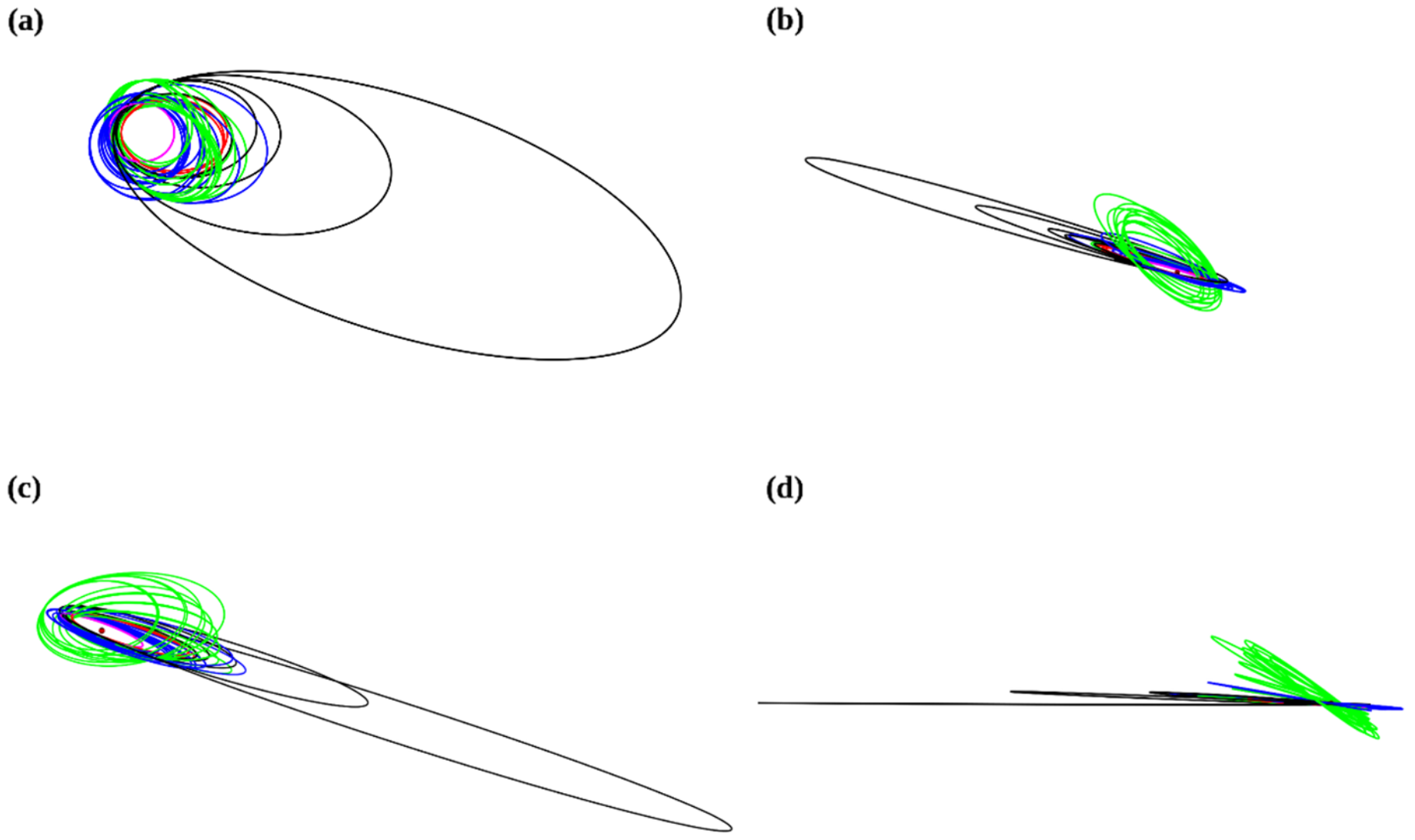

**Fig. 3** View of the trajectory in the Jupiter system. The colour coding illustrates the mission phase (see Table 4 and Boutonnet et al. 2024): energy reduction phase (black), Europa phase (red), high-latitude phase (green), transfer to Ganymede (blue), Ganymede orbital phase (pink). Panel a) displays the trajectory view from the top, panels b), c) and d) shows a different viewing angle to better see the differences in inclination

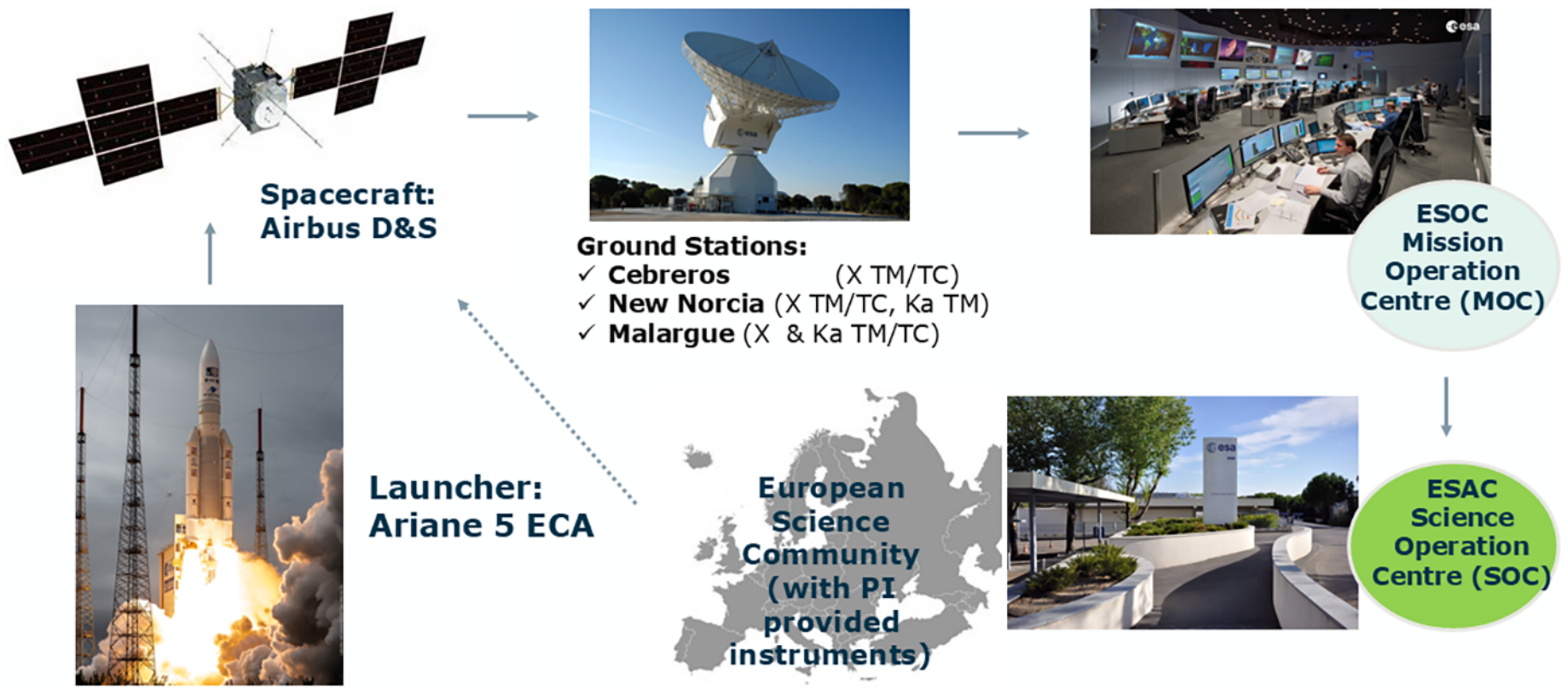

**Fig. 4** Overall mission organisation, or mission elements (courtesy of ESA). Support from the NASA Deep Space Network is foreseen as part of a Memorandum of Understanding between ESA and NASA

in Fig. 4. Among others, the key aspect was the control of the schedule of each element to avoid any delays in one element that would impact the timing of another, with associated cost overrun. The development of the spacecraft and the associated instruments were the driver, but the procurement of the launch rocket and the preparation of the launch pad, the development of the operation and science ground segment software and infrastructure had to follow in synchronisation. This considerable endeavour required a large organisation effort at many levels to coordinate the scientific, industrial, and operation teams involved. Such coordination implied the implementation of different types of interfaces and relations:

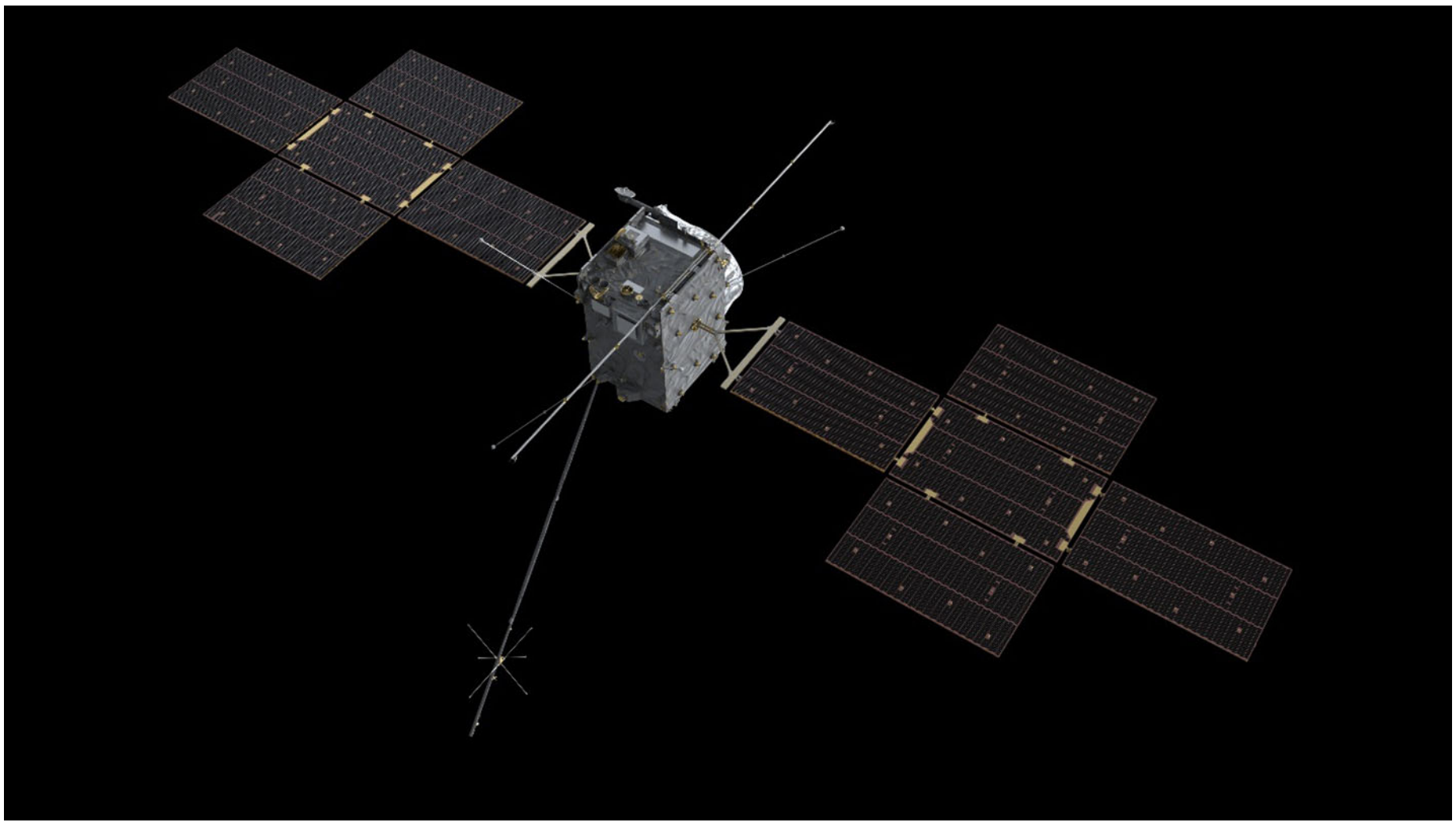

**Fig. 5** The JUICE spacecraft (6 tonnes at launch), showing its cuboidal body, twin cross-shaped solar-panel wings, and extended booms and antennas

contractual as far as the industrial elements (spacecraft and launch system) were considered and based on cooperation for the instruments and operational aspects.

The overall cost at completion of the mission is around 1.6 B€ (billion) euro, roughly distributed as follows: spacecraft development 35%, launch system procurement 11%, instruments development 37%, ground segment and project team 17%. The cost includes the in-flight operation up to the end of the nominal mission planned end of 2035.

## 5.1 The Spacecraft

JUICE is the largest spacecraft (see Fig. 5) that ESA has ever sent in the outer solar system with a mass at launch of about six tonnes (Sarri et al., this collection). This is because travelling to Jupiter and achieving the final orbits, first around Jupiter and then around Ganymede at the end of the mission, requires a very large amount of propellant: 3.6 tonnes. It will take approximately eight years to reach Jupiter. This long journey is required due to the high launch mass of the spacecraft attaining the limits of transfer to Jupiter even by the most powerful version of the Ariane 5 rocket. Therefore, the spacecraft uses a sequence of four gravity assist manoeuvres (Moon and Earth, Venus, Earth and again Earth). JUICE will experience a very large range of thermal conditions during its journey: from the extremely hot temperatures during the Venus fly-by and at Sun closest approach to the extremely cold ones at Jupiter. It will have to survive the harsh radiation environment of space and especially within the strong magnetic field of Jupiter. This places stressful conditions on all the electronic units. Since Jupiter is more than five times further away from the Sun than the Earth, the solar flux there is only 1/25th of that on Earth. This means that, during its nominal operation, the available power will be limited to about 800 W even with the very large solar arrays. Each electronic unit was designed to favour power consumption minimisation; strict power requirements were placed on all instruments. The key design drivers that eventually shaped the spacecraft are summarised in Table 6.

**Table 6** Key design drivers of the JUICE mission (From Sarri et al. 2026, this collection)

| The driver | The problem | The solution |
|---|---|---|
| Very little power away from the Sun | ∘ Solar flux at Jupiter around 50 W/m$^2$<br>∘ Solar cells efficiency at cold | → Large solar array size: 85 m$^2$ with customised LILT (Low Intensity Low Temperature) solar cells |
| High and low solar flux | ∘ Very low at Jupiter → ∼ 50 W/m$^2$<br>∘ Very high at Sun closest approach → 3320 W/m$^2$ | → High performing multi-layer insulation<br>→ High gain antenna (with special white paint) as thermal shield |
| Harsh radiation environment | ∘ High energy particles<br>∘ Total Ionising Dose → 1 GRad | → Highly radiation tolerant components<br>→ Local and global shielding<br>→ Two dedicated protective vaults |
| Extreme instrument's sensitivity to electromagnetic disturbances | ∘ Instruments measuring fields and particles<br>∘ Spacecraft generated magnetic fields must be less than 1 nT (50,000 times less than Earth magnetic field) | → Special design of<br>■ Solar array<br>■ Power system<br>■ Reaction wheels<br>■ Shielding and grounding<br>■ Magnetic cleanliness<br>operations during integration and test |
| Complex mission (navigation and autonomy) | ∘ 4 gravity assists during the cruise to Jupiter<br>∘ 35 fly-bys of the Galilean moons at Jupiter; down to 12 days between consecutive flybys<br>∘ Signal turnaround time at Jupiter → 1.5 h | → Spacecraft highly autonomous<br>→ Complex Failure Detection, Isolation and Recovery (FDIR) |
| Large amount of propellant necessary | ∘ Deep space and orbit correction manoeuvres<br>∘ Jupiter orbit insertion<br>∘ Ganymede orbit insertion | → 3.6 tonnes of propellant<br>→ Two large tanks<br>→ Strong, but light, structure |

Figure 6 shows the final configuration of the spacecraft viewed from the top (the Nadir side) with the identification of the instruments. Figure 7 shows the spacecraft after its arrival at the launch site. The two main antennas, High Gain and Medium Gain, are under the protective cover. Particularly noticeable is the large size of the High Gain Antenna, which serves as the primary communication link at Jupiter. Its diameter of 2.5 m is driven by the required average science downlink rate of 1.4 Gbit per day. Because observation and data transmission to Earth do not occur simultaneously, a high-capacity Solid-State Mass Memory composed of two redundant modules is implemented on JUICE. This system provides a minimum capacity of 1 Tbit throughout the mission, which is sufficient to store the data collected by the payload during simultaneous instrument operations, including the most demanding Europa flyby. On the top and on the bottom of the spacecraft are the folded solar arrays. The red element on the top is protective cover of the solar array, not yet removed. Tables 7 and 8 show the spacecraft' s main parameters and the detailed mass budget.

The spacecraft was built by a large industrial consortium led by Airbus Defence and Space (ADS) also called the Prime Contractor, and under the overall programme management of the European Space Agency (ESA). The industrial consortium (Fig. 8) was structured in three levels: the upper was the prime contractor supported by few selected companies altogether forming the "core team". The second level was made by six companies in charge of key subsystems of the spacecraft (e.g. the solar array). The third level was made by several companies providing equipment (e.g. the star trackers or the on-board computer). Overall, 109 companies were part of the JUICE industrial consortium, 144 contracts were awarded and managed.

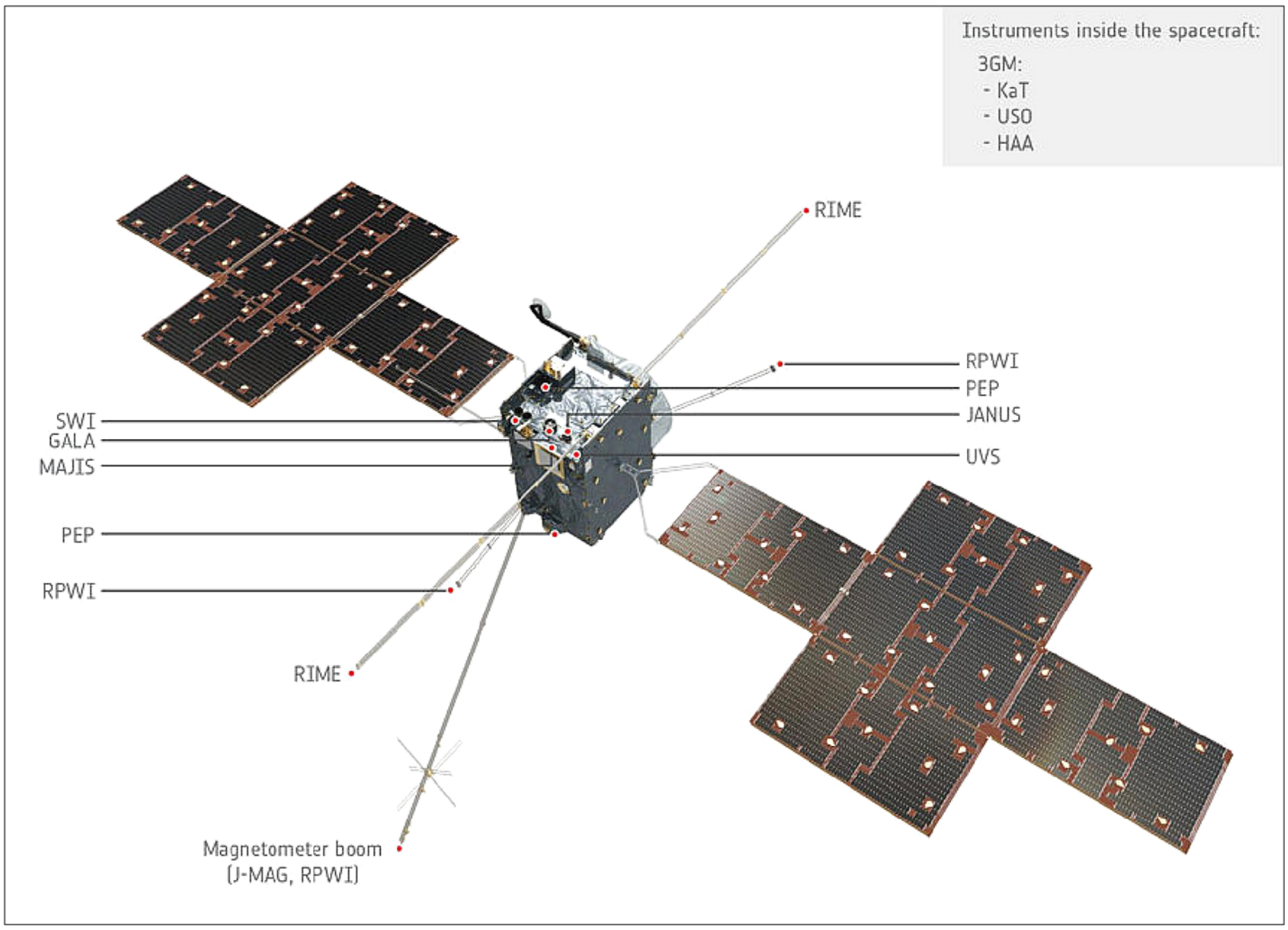

**Fig. 6** Spacecraft configuration showing the position of the ten instruments (courtesy of ADS)

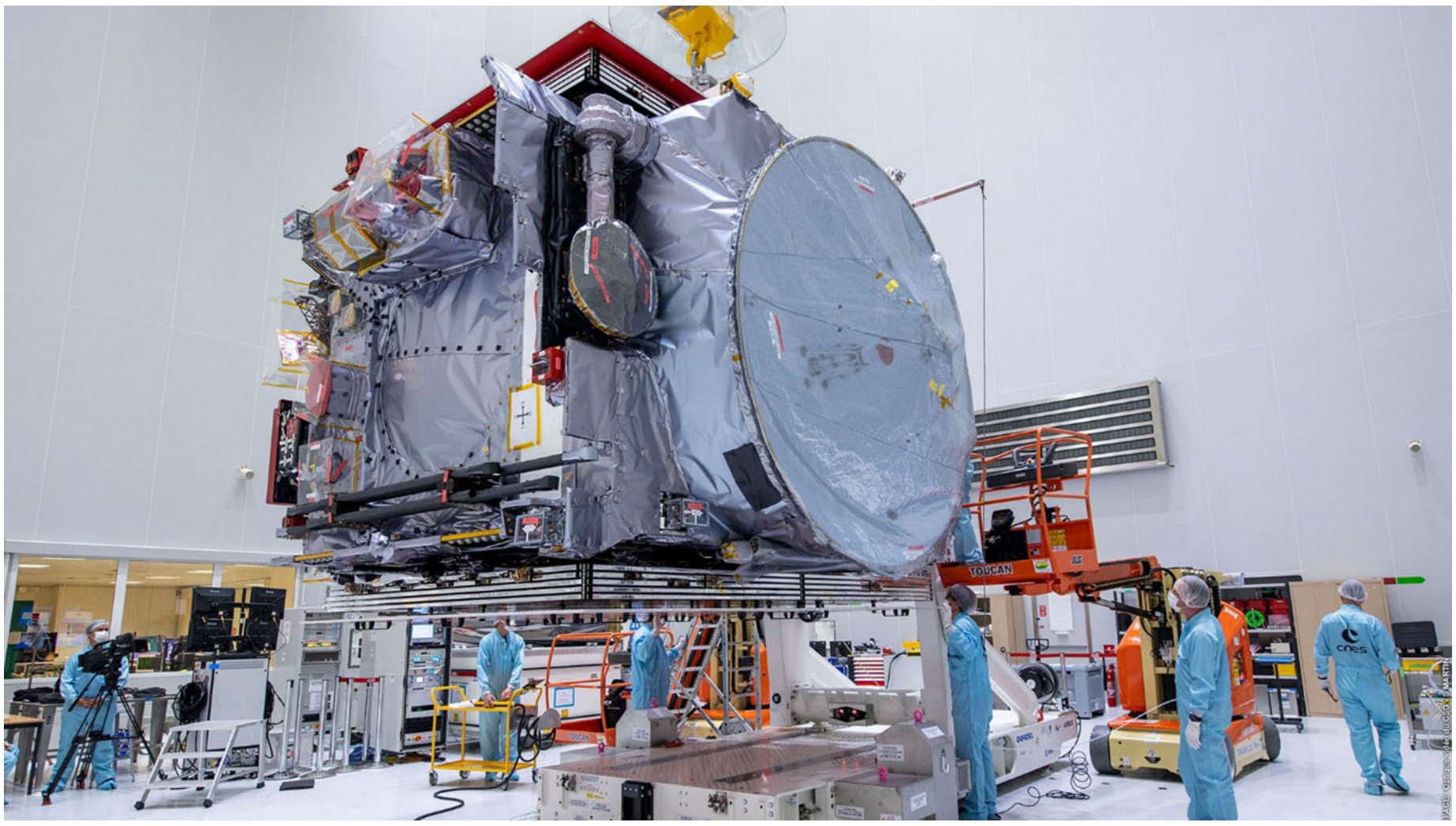

**Fig. 7** The JUICE spacecraft, its solar arrays folded up at top and bottom, in the clean room at the Kourou launch site (courtesy of ESA). High-gain and medium-gain antennas are the large circular elements in foreground

**Table 7** JUICE spacecraft main parameters

| | |
|---|---|
| **Dimensions** | |
| S/C envelope (solar arrays stowed): | 4.4 m × 4.1 m × 2.9 m |
| S/C envelope (solar arrays deployed): | 27 m × 7.7 m |
| **Mass** | |
| Total launch mass: | 6051 kg |
| S/C dry mass with instruments | 2299 kg |
| Instruments: | 250 kg |
| Propellant and pressurant: | 3657 kg |
| S/C to launcher adaptor: | 95 kg |
| **Solar Array** | |
| Surface: | 85 $m^2$ |
| Number of solar cells: | 23,560 (triple junction Gallium Arsenide) |
| Output power end of life: | 780 W @ Jupiter |
| **Batteries** | 6 kWh (5 modules) |
| **Propulsion subsystem** | |
| Thrust | 1 Main Engine (425 N)<br>8 × 22 N thrusters, nominal and redundant |
| Tanks | 12 × 10 N thrusters, nominal and redundant |
| Propellant fuel | 2 × 1630 l propellant tanks |
| Propellant oxidiser | 3 × 90 l pressurant tanks |
| Pressurant | MMH (Monomethylhydrazine)<br>NTO (Nitrogen Tetroxide)<br>Helium |
| **Communication subsystem** | |
| 2 × Deep Space Transponders | X-band uplink<br>X-band and Ka-band downlink |
| 1 High Gain Antenna | X- and Ka-band |
| 1 Medium Gain Antenna | X- and Ka-band |
| 2 × Low Gain Antennas | X-band, for early orbit phase and emergency operations |
| **Attitude Control and Navigation** | 2 × Sun Sensors<br>3 × Star Trackers<br>2 × Inertial Measurement Sensors, with gyros and accelerometers<br>4 × Reaction Wheels<br>2 × Navigation Cameras |
| **Appendages** | |
| Spacecraft | 2 × solar arrays (10 panels)<br>1 × deployable and steerable Medium Gain Antenna |
| Instruments | 1 Magnetometer boom (with 1 deployable antenna on it)<br>1 RIME antenna (6 segments)<br>4 Langmuir probes<br>Deployable covers |
| **Reliability and operational lifetime** | 0.88 after 13 years nominal life |

**Table 8** JUICE spacecraft detailed dry mass breakdown

| Spacecraft Subsystem | Mass (kg) |
|---|---|
| Structure | 656 |
| Radiation shielding | 86 |
| Harness | 170 |
| Thermal control | 135 |
| Power | 428 |
| Propulsion | 281 |
| Communications | 100 |
| Attitude Control and Navigation | 105 |
| Data Handling | 31 |
| Monitoring and calibration systems | 11 |
| Instruments | 250 |
| Radiation Monitor | 5 |
| Magnetometer boom | 34 |
| RIME antenna | 7 |
| **Total spacecraft dry mass** | **2299** |

## 5.2 JUICE Instrumentation and Scientific Consortium

The payload consists of 10 state-of-the-art instruments, one experiment that uses the spacecraft telecommunication system with ground-based elements, plus a radiation monitor. The instruments are divided into three main packages: remote-sensing, geophysics and in-situ. The remote sensing package includes imaging (JANUS) and spectral-imaging capabilities from the ultraviolet to the sub-millimetre wavelengths (MAJIS, UVS, SWI). The geophysical package consists of a laser altimeter (GALA) and a radar sounder (RIME) for exploring the surface and subsurface of the moons, and a radio science experiment (3GM) to probe the atmospheres of Jupiter and its satellites and to perform measurements of the gravity fields. An experiment (PRIDE) using ground based Very Long Baseline Interferometry (VLBI) will provide supporting tracking data with the focus of improving the ephemerides of the Jovian system. The in-situ package comprises a full suite to study plasma and neutral gas environments (PEP) with remote sensing capabilities of energetic neutrals, a magnetometer (J-MAG) and a radio and plasma wave instrument (RPWI), including electromagnetic fields sensors and a Langmuir probe. Table 9 gives a synthetic overview of the JUICE payload while Fig. 9 details their location on the spacecraft. Figure 10 illustrates the contributions from ESA countries and US PI institutes. Table 10 lists the "facility instruments" onboard JUICE, provided and funded by ESA. The following subsections provide a short overview of each instrument.

### 5.2.1 Jovis, Amorum Ac Natorum Undique Scrutator (JANUS)

The JANUS instrument (Palumbo et al. 2025), the multispectral camera enabling imaging in the 340–1080 nm wavelength range, has been designed for optimum performances when imaging Ganymede, Europa, Callisto, Io, small and irregular moons, Jupiter and the rings. The spatial resolution for the three icy Galilean satellites will range from kilometres to a few metres per pixel. At Ganymede it will be 7.5 m/pixel at 500 km distance and 3 m/pixel at 200 km altitude. For the other targets the resolution will be several km per pixel. 13 filters can guarantee good spectral coverage with bandwidths from several tens of nm down to 10 nm. The JANUS resolution, in its spectral bands, will provide unprecedented characterisation of endogenic and exogenic geological processes that shaped the Icy satellites surfaces, monitor Io's volcanic activity, investigate physical and dynamical properties of small satellites and rings, as well as the dynamics of Jupiter atmosphere.

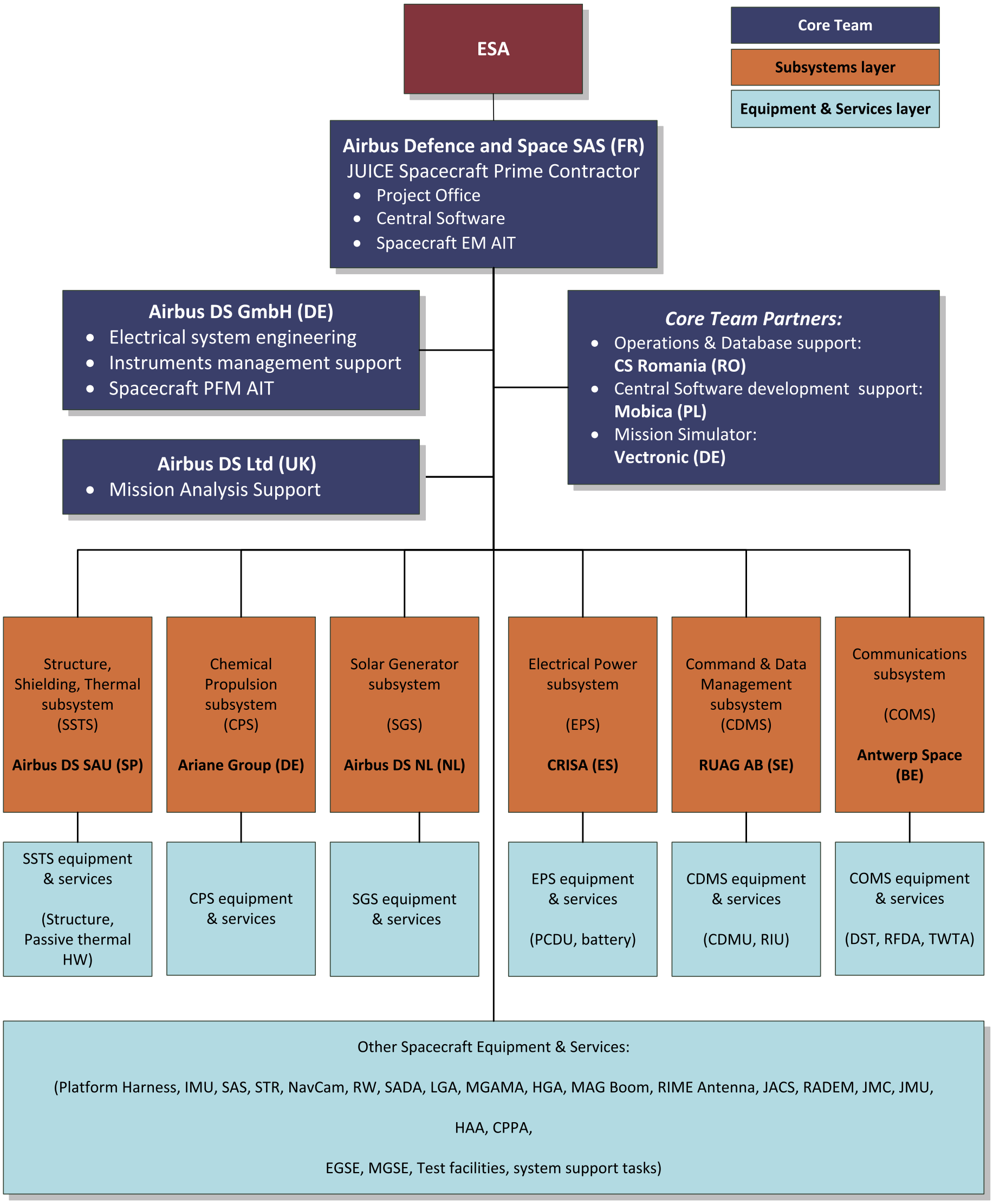

**Fig. 8** JUICE spacecraft industrial set up (courtesy of ADS)

**Table 9** Summary of JUICE's scientific instrumentation

| Instrument name (acronym) | Principal Investigator | Science objectives and measurements | Mass (kg) | Typical power (W) |
|---|---|---|---|---|
| Jovis Amorum ac Natorum Undique Scrutator (JANUS) | P. Palumbo, INAF, Rome, Italy | A multispectral camera (13 filters in the visible wavelength range) to image all targets of the mission. JANUS will study geological processes that shaped the icy satellites surfaces, monitor Io's volcanic activity, investigate properties of small satellites and rings, and the dynamics of Jupiter atmosphere. | 24.5 | 20 |
| Moons and Jupiter Imaging Spectrometer (MAJIS) | F. Poulet, Institut d'Astrophysique Spatiale Orsay, FRANCE | An imaging spectrometer operating across the 0.50–5.55 µm visible and near-infrared range to characterise the composition of the icy moons' surfaces and exospheres and observe Jupiter's cloud structures, atmospheric constituents, auroras, ring system, and minor moons. The measurement strategy includes nadir and off-nadir viewing, limb-pointing observations, and stellar occultations. | 51 | 25 |
| Ultraviolet Spectrograph (UVS) | K. Retherford, SwRI, San Antonio, USA | An ultraviolet imaging spectrograph operating over the 50–204 nm range to characterise the composition and dynamics of the icy moons' exospheres, study the auroras of Ganymede and Jupiter, trend emissions from Io's atmosphere and its Io Plasma Torus, and probe the composition and structure of Jupiter's upper atmosphere. UVS will also attempt to detect possible plumes at Europa. Data collection will take place in nadir geometry, limb-pointing observations, distant scans, Jupiter transits, mutual events, and during both solar and stellar occultation opportunities. | 19 | 8 |
| Sub-millimetre Wave Instrument (SWI) | P. Hartogh, Max-Planck-Institut für Sonnensystem-forschung, Göttingen, Germany | A heterodyne spectrometer operating in two separate GHz-frequency bands (1080–1275 GHz and 530–601 GHz) to probe the temperature structure, composition, and dynamics of Jupiter's atmosphere, as well as the exospheres and surfaces of the icy moons. | 19 | 55-65 |

**Table 9** (*Continued*)

| Instrument name (acronym) | Principal Investigator | Science objectives and measurements | Mass (kg) | Typical power (W) |
|---|---|---|---|---|
| Ganymede Laser Altimeter (GALA) | H. Hussmann, DLR, Germany | Laser altimeter (wavelength of 1064 nm) to study the tidal deformation of Ganymede, the morphology and topography of the surfaces of the icy moons, and their roughness. | 23 | 55 |
| Radar for Icy Moon Exploration (RIME) | L. Bruzzone, Università degli Studi di Trento, Italy | A 9-MHz ice penetrating radar using a 16 m dipole antenna to probe the subsurface of the icy moons to depths down to nine kilometres, enabling the detection of internal structures, potential water pockets, and the ice-water interface within the ice shell. | 12 | 15-20 |
| Gravity and Geophysics of Jupiter and Galilean Moons (3GM) | L. Iess, Università di Roma "La Sapienza", Italy | A radio science package comprising a Ka-band transponder and an ultra stable oscillator, complemented with a high accuracy accelerometer (HAA) as a facility instrument (see Table 10), to study the gravity fields of the icy moons, investigate the dynamics (ephemerides) and interiors of these moons, as well as the structure of the neutral atmospheres and ionospheres of Jupiter and its moons. 3GM will also study the Io torus and the ring system. | 6 (HAA not included) | 40 (HAA 12 W not included) |
| JUICE magnetometer (J-MAG) | M. Dougherty, Imperial College London, United Kingdom | A suite of two fluxgate and one scalar magnetometer to characterise the interactions between the Jovian magnetosphere and the icy moons, study the internal Ganymede's magnetic field, and characterise the subsurface oceans of Ganymede and Callisto. | 6 | 7-17 |
| Particle Environment Package (PEP) | S. Barabash, Swedish Institute of Space Physics, Kiruna, Sweden | A package of six instruments to characterise the plasma environment in the Jovian system. It will measure density and fluxes of positive and negative ions, electrons, exospheric neutral gas, thermal plasma, and energetic neutral atoms in the energy range from $<0.001$ eV to $>1$ MeV with full angular coverage. The composition of the moons' atmospheres will be measured with a resolving power of more than 1000. | 50 | 60-80 |

**Table 9** (*Continued*)

| Instrument name (acronym) | Principal Investigator | Science objectives and measurements | Mass (kg) | Typical power (W) |
|---|---|---|---|---|
| Radio and Plasma Wave Investigation (RPWI) | J.-E. Wahlund, Swedish Institute of Space Physics, Uppsala, Sweden | A radio and plasma wave package to characterise electromagnetic fields and plasma environment of Jupiter and its icy moons. It will use Langmuir probes to measure DC electric field vectors and to characterise thermal plasma; medium- and high-frequency receivers, dipole antennas and magnetometers to measure electric and magnetic fields in radio emission in the frequency range 80 kHz–45 MHz | 17 | 14-21 |
| Planetary Radio Interferometry and Doppler Experiment (PRIDE) | G. Cimo, Joint Institute for VLBI ERIC, The Netherlands | Using ground-based radio telescopes, the investigation aims to study the ephemerides, and atmospheric properties of the Jovian moons through complementary radio science observables (VLBI and Doppler) | 0 | 0 |

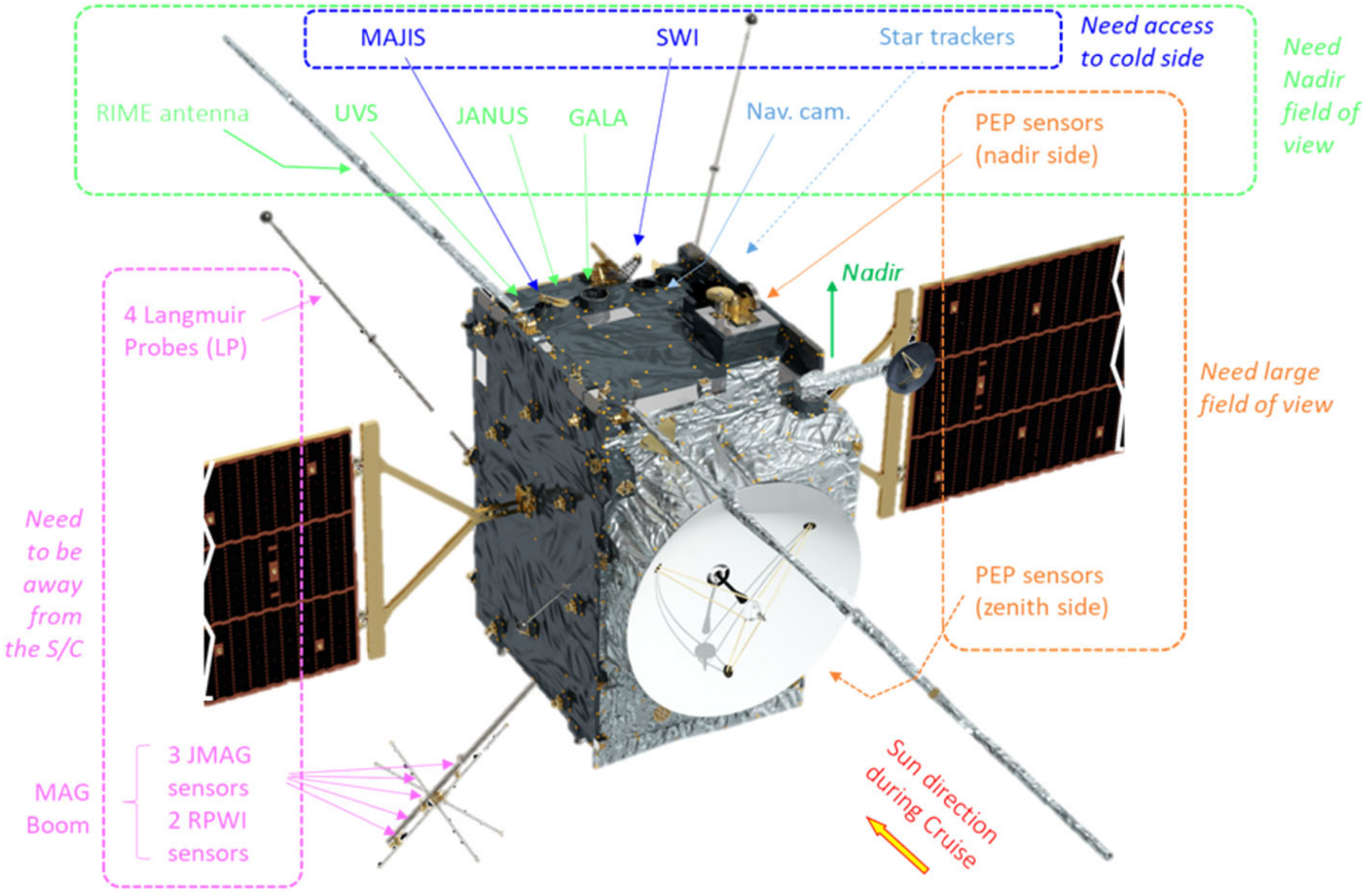

**Fig. 9** Instrument accommodation on the spacecraft (courtesy of ESA and ADS. From Sarri et al. 2026, this collection)

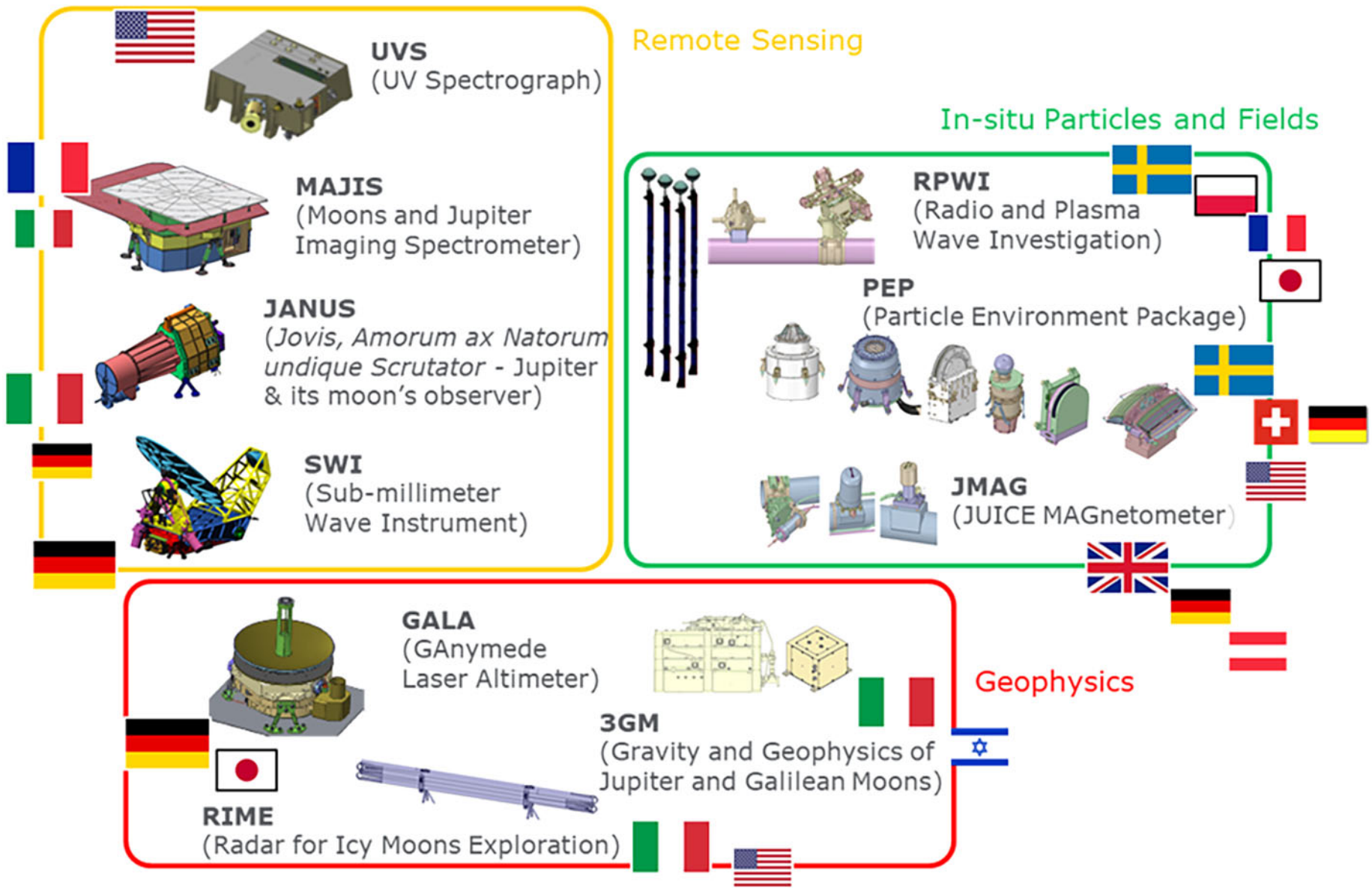

**Fig. 10** Leading countries contributing to the instruments on board JUICE (courtesy of ESA)

**Table 10** Summary of JUICE's facility instruments

| Facility instrument name (acronym) | Objectives |
|---|---|
| RADiation–hard Electron Monitor (RADEM) | Monitoring of the radiation environment by measurement of electrons, protons and heavy ions. |
| High Accuracy Accelerometer (HAA) | Measurement of non-gravitational accelerations to enable the determination of the spacecraft's centre of mass position. It supports the 3GM instrument investigations. |
| NAVigation CAMera (NAVCAM) | Two redundant and independent cameras to be used for navigation. |
| | Images are also used for public relations purpose. Images can be used for context, science, astrometry. |
| JUICE Monitoring Cameras (JMC) | Two low-rate colour cameras to provide video footage of the post-launch deployment of the solar arrays and appendages. Images are also used for public relations purpose. Images can be used for context. |

### 5.2.2 Moons And Jupiter Imaging Spectrometer (MAJIS)

The MAJIS imaging spectrometer (Poulet et al. 2024), is an imaging spectrometer operating in the visible and near-infrared spectral range. With nominal sampling the spectral resolution will be better that 4 nm/band for the visible/near infrared channel (0.50 to 2.35 μm), and better than 7 nm/band for the infrared channel (2.25 μm to 5.54 μm). It will be possible to perform spectral oversampling over selected spectral ranges representing up to 20% of the total range of each channel. The scientific field of view (FOV) is 3.4° along the slit, ±1.7° with respect to the optical boresight with an IFOV (Instaneous FOV) of 150 μrad (31 arcsec, 400 pixels over the FOV). As already demonstrated by the past and present operative planetary space missions, an imaging spectrometer of this type can span a wide range of scientific objectives, from the surface through the atmosphere and exosphere. MAJIS is then perfectly suitable for a comprehensive study of the Galilean icy satellites, with particular emphasis on Ganymede, the Jupiter atmosphere, including its aurorae and the spectral characterisation of the whole Jupiter system. The accurate measurement of radiance from the different targets, in some case particularly faint due to strong absorption features, requires a very sensitive cryogenic instrument operating in a severe radiation environment. Its passive cooling system without mechanical cryocoolers ensures MAJIS robustness for the long-life mission that JUICE requires.

### 5.2.3 Ultraviolet Spectrograph (UVS)

UVS (Retherford et al. 2027, this collection) is the fifth in a series of successful ultraviolet imaging spectrographs (Rosetta-Alice, New Horizons Pluto-Alice, LRO-LAMP, and Europa Clipper UVS) and is largely based on the most recent of these, the Juno-UVS. The goals of the UVS investigation are to explore the atmospheres, plasma interactions, and surfaces of the Galilean satellites; to determine the dynamics, chemistry, and vertical structure of Jupiter's upper atmosphere, from equator to pole; and to investigate the Jupiter-Io connection by quantifying energy and mass flow in the Io atmosphere, neutral clouds, and torus. The instrument observes photons in the wavelength 50 to 204 nm range, at moderate spectral and spatial resolution along a 7.5° opening slit. A main entrance "airglow port" is used for most observations (e.g., airglow, aurora, surface mapping, and stellar occultations), while a separate "solar port" allows for solar occultations. Another aperture door, with a small square aperture through the centre, is used as a "high-spatial-resolution port" for detailed observations of bright targets. Time-tagging and programmable spectral imaging allow for observational flexibility and optimal data management.

### 5.2.4 Sub-Mm Wave Instrument (SWI)

The JUICE Submillimetre Wave Instrument (SWI) is a passive heterodyne microwave spectrometer sensitive for radiation in the frequency band of 530–625 GHz for the channel one (Band#1) and 1080–1275 GHz (Band#2) for channel two (Kotiranta et al. 2018; Hartogh et al. 2027, this collection; Kotiranta et al. 2026). The radiation is received through a quasi-optical off-axis telescope with 29 cm aperture diameter, providing a spatial resolution of 2 mrad (FWHM) at 600 GHz. This telescope is equipped with a two-axis scan mechanism to allow scanning around the nadir viewing direction in a range of ±72° inside the spacecraft's orbital plane about Jupiter and ±4.3° perpendicular to the plane (north-south when in Jupiter orbit). The instrument will provide a detailed characterisation of the dynamics, and composition of Jupiter's stratosphere, and its coupling with the lower and upper atmosphere. It will characterise the tenuous atmospheres and exospheres of the icy satellites, enabling the determination of their sources and sinks, interactions with the Jovian magnetosphere and measure their surface and subsurface properties. It will determine key isotope ratios in Jupiter's and satellite atmospheres, constraining the origin and evolution of the Jovian system. Two independent double sideband receivers are used to obtain simultaneous observing capability for two different frequencies bands. For radiometric calibration, the SWI instrument uses an internal blackbody as hot reference load and the cold sky as external cold reference load.

### 5.2.5 Ganymede Laser Altimeter (GALA)

GALA (Hussmann et al. 2019, 2025; Kimura et al. 2019; Enya et al. 2022) is a Nd:YAG laser altimeter, in which neodymium ions are doped into an yttrium aluminium garnet crystal. The instrument operates at an emission wavelength of 1064 nm with a pulse repetition frequency nominally of 30 Hz, of up to 48 Hz. GALA will acquire multiple topographic profiles during flybys of Europa and Callisto and will generate a high-resolution global shape model of Ganymede from circular orbits at altitudes of 500 km and 200 km. By tracking Ganymede's diurnal tidal deformation, the instrument will test for the presence of a subsurface ocean and constrain the thickness of the overlying ice shell. To achieve the spatial resolution required for geological characterisation and surface roughness analysis, the GALA footprint diameter is on the order of 50 m at the 500 km orbital altitude. Furthermore, to obtain extended laser ground tracks for determining the global figure of Europa and Callisto and to characterize surface features from precise height profiles during flybys, the receiver can detect return signals from distances of up to approximately 1100–1600 km from the target body. The maximum detectable range depends on the surface albedo of the respective moon.

### 5.2.6 Radar for Icy Moons Exploration (RIME)

The Radar for Icy Moons Exploration (RIME) is a Low frequency radar designed to investigate the subsurface of icy moons (Bruzzone et al. 2026, this collection). It is a monostatic, high-frequency radar operating at 9 MHz and transmitting linearly frequency-modulated pulses, with a maximum theoretical bandwidth of 3 MHz (maximum practical bandwidth: 2.8 MHz). RIME is a pulsed radar system transmitting programmable radio-frequency (RF) pulses at repetition frequencies between 100 Hz and 1 kHz. The nominal pulse width is 100 μs, with can be adjusted in the range of 50 to 250 μs. Owing to its low operating frequency, significant penetration depths, up to several km, can be achieved in low-attenuation media such as water ice. RIME radiates through a dipole antenna, that was not provided by the instrument team itself but provided and procured by the spacecraft prime contractor. RIME consists of three main units: the Receiver, the Transmitter, and the Matching Network, which are electrically interconnected. Control functions, signal generation, and receive-chain processing are implemented within the Receiver unit, while the Transmitter provides power amplification of the outgoing signal. The Matching Network is a passive unit that ensures impedance matching between the highly reactive and frequency-dependent antenna load and the transmitter interface. Digital processing of the received signal is performed in hardware, while a LEON-based processor is employed for instrument control functions.

### 5.2.7 Gravity and Geophysics of Jupiter and Galilean Moons (3GM)

The scientific objectives of 3GM can be divided into gravity science, radio occultation experiments, and bistatic radar investigations (Iess 2027, this collection; Cappuccio et al. 2025). The radio science elements of the instrument are essential for a comprehensive understanding of Jupiter's moons. The gravity measurements, complemented by accurate determination of orbital parameters, rotation state, and precise shape reconstruction in conjunction with the laser altimeter, will provide an unprecedented view of the bodies' differentiation state, internal dynamics, and ultimately their evolution. The radio-occultation experiment will measure the physical properties of Jupiter's atmosphere, as well as those of its moons. Bistatic radar observations will constrain the dielectric permittivity of Ganymede's surface material. In addition to the Ka-band Transponder (KaT) and the Ultra-Stable Oscillator (USO), which are the core elements of the instrument, the system includes a ground-based data processing facility for scientific analysis and an onboard HAA. The KaT provides a two-way Ka-band link with the ground station via the spacecraft HGA, when pointed toward Earth, enabling accurate range and range-rate measurements. During periods when attitude constraints require different spacecraft pointing (e.g., due to remote sensing instrument operations, both in orbital phases and flybys), the KaT uses the steerable MGA to collect tracking data in the vicinity of Jupiter's moons. All 3GM data are processed on the ground, together with accelerometer data. This enables reconstruction of the spacecraft's trajectory and estimation of the scientific parameters of interest. The USO provides a highly stable frequency reference for the radio occultation experiment, as studies of the ionospheres and atmospheres require one-way downlink measurements in both X- and Ka-band.

### 5.2.8 JUICE Magnetometer (J-MAG)

The J-MAG magnetometer (Dougherty et al. 2027, this collection; Amtmann et al. 2024) was primarily designed to measure Ganymede's induced magnetic response, thereby enabling characterisation of its subsurface ocean, including estimates of its depth and salinity. The instrument will also investigate Ganymede's complex magnetic environment and perform additional measurements at Callisto, Europa, and throughout Jupiter's magnetosphere. J-MAG measures the three components of the local magnetic field vector in the vicinity of the spacecraft. The instrument comprises two fluxgate sensors and one scalar sensor. All three sensors are mounted on a 10.6-m boom, provided by the industrial prime, and are connected to an electronics unit housed within the spacecraft body. Each fluxgate is a tri-axial sensor that measures the three components of the magnetic field, while the scalar sensor measures the magnitude of field strength. J-MAG is required to measure the local magnetic field over timescales ranging from days to frequencies of several hertz, with a relative precision of a few pT and an absolute accuracy, including spacecraft-generated magnetic effects, of 0.2 nT or better.

### 5.2.9 Particle Environment Package (PEP)

The Particle Environment Package was designed to investigate particle populations in the Jovian magnetosphere and in the vicinity of the Galilean moons through measurements of neutral gas, thermal plasma, positive and negative ions, electrons, and energetic neutral atoms over an energy range from 1 eV to 5 MeV (Barabash et al. 2027, this collection). PEP comprises six instruments distributed on the spacecraft to ensure optimal field-of-view coverage. JEI, JNA, and NIM, together with their associated electronics, are mounted on a common panel on the nadir side. JENI is also located on the nadir side, at a separate position, while JoEE and JDC are mounted on the zenith side. JEI (Jovian Electron and Ion Analyser) measures thermal and suprathermal electrons and ions without mass discrimination along the JUICE trajectory. It provides full hemispherical coverage in the energy range of 1 eV to 20 keV and limited field-of-view coverage from 20 keV to 50 keV. JNA (Jovian Neutrals Analyser) is a surface-interaction energetic neutral atom instrument (Wurz 2000). Incident neutrals are ionised on a conversion surface, energy-analysed, and post-accelerated into an integrated time-of-flight section. The sensor and electronics are accommodated within a single compact unit. NIM (Neutral Gas and Ion Mass Spectrometer) is a time-of-flight mass spectrometer designed for neutral gas and ion composition measurements. NIM consists of a titanium tube integrating the ion source and detector at one end and a reflectron at the other. A heavily shielded microchannel plate detector (Lasi et al. 2017) mitigates noise induced by Jupiter's radiation environment. JDC (Jovian Plasma Dynamics and Composition) is a reflectron-type time-of-flight ion mass spectrometer providing full hemispherical coverage and mass-resolved measurements of positive and negative ions. JENI (Jovian Energetic Neutrals and Ions) combines an energetic ion sensor with an energetic neutral atom imager. It employs serrated magnesium deflection plates to separate charged particles via an applied electric field prior to detection. JoEE (Jovian Energetic Electrons) is a compact medium- to high-energy electron spectrometer in which electrons are guided by a self-closing permanent magnetic field onto segmented solid-state detectors.

### 5.2.10 Radio and Plasma Wave Investigation (RPWI)

The Radio and Plasma Wave Investigation (Wahlund et al. 2025, this collection) provides an elaborated suite of electromagnetic fields and cold plasma instruments, where several different types of sensors will sample the thermal plasma, DC electric fields, electric and magnetic signals from radio, plasma waves and micrometeorite impacts. RPWI focuses, apart from cold plasma studies, on the understanding of how, through electrodynamic and electromagnetic coupling, the momentum and energy transfer occur in the surrounding space environments and with the icy Galilean moons. In Ganymede orbit, in-situ measurements by RPWI of the electric coupling between any sub-surface ocean, the ionosphere, and the magnetosphere will provide constraints on the physical characteristics of a sub-surface ocean. The instrument consists of three groups of instruments, with 10 sensors in total. Four Langmuir probes for thermal plasma and electric field measurements from DC to 1.5 MHz, three axes search coil magnetometer for magnetic fields measurement to measure the vector magnetic and electric fields simultaneously below 20 kHz, three dipole antennas for radio measurements in the frequency range 80 kHz–45 MHz to fully characterise radio emissions, making it possible to determine source locations and polarisation of radio emissions from the aurora, Jupiter's magnetosphere, Ganymede's magnetosphere, and characterise their variability.

### 5.2.11 PRIDE

The Planetary Radio Interferometry and Doppler Experiment (Gurvits et al. 2023, this collection; Pallichadath et al. 2025) is an investigation that utilises the spacecraft's communication system in combination with Earth-based radio telescopes to record the JUICE spacecraft signal. By determining the spacecraft's lateral position through very long baseline interferometry and measuring the radial Doppler shift at all observing stations, PRIDE data will improve planetary and satellite ephemerides (Fayolle et al. 2024), support radio occultation experiments, and probe interplanetary plasma conditions. The PRIDE dataset will also help validate and, where possible, refine estimates of the Galilean satellites' gravity fields and the spacecraft orbit.

### 5.2.12 Facility Instruments

JUICE also embarks a few facility instruments, which are subsystems provided under ESA responsibility: a high-accuracy accelerometer (HAA), a radiation monitor (RADEM), two navigation cameras (NAVCAM) and two Jupiter monitoring cameras (JMC). The HAA measures non-gravitational accelerations along three orthogonal reference axes to determine the spacecraft's centre-of-mass position with the precision needed for the 3GM gravity-field experiment. HAA is used to calibrate non-gravitational perturbations affecting Doppler measurements, such as those caused by propellant sloshing in the spacecraft tanks. The data processing and analysis is under the responsibility of the instrument team. Calibration information can be found in Astrua et al. (2023). Activated during the deployment of appendages shortly after launch, HAA detected the main vibration modes, enabling a comprehensive analysis of the spacecraft's dynamic response (De Filippis et al. 2024). The HAA can also be used to study vibrations excited by instrument moving parts and outgassing events (De Filippis et al. 2026). RADEM (Hajdas et al. 2025, this collection; Rodriguez-Garcia et al. 2025; Pinto et al. 2026b) is a permanently operating radiation monitor providing data on energetic particle fluxes and their energy spectra. In addition of being a unit with the main goal of monitoring the radiation environment, RADEM data can be used for science purposes, extending the energy range covered by the PEP instrument. RADEM consists of three detector subunits: the Electron Detector Head, the Proton and Heavy Ion Detector Head, and the Directional Detector Head. RADEM measures electrons in the 0.3–40 MeV energy range, protons in the 5–250 MeV energy range, and heavy ions within the Linear Energy Transfer range from 0.1 to 10 MeV cm/mg. The angular coverage of incoming radiation is up to about 35% of the sky.

The two NAVCAM (Gorog et al. 2019) are cameras used for navigation purposes. Their images can be used for scientific purposes as well. The two JMC (Adamiec et al. 2021) were designed to provide images of the deployment of the solar panels and appendages shortly after launch and can be used in cruise and during the nominal mission.

#### 5.2.13 Science Working Team Organisation

As documented in the Science Management Plan, the JUICE Science Working Team (SWT) serves as the central body overseeing and coordinating the mission's scientific activities. Chaired by the Project Scientist, it formally includes the eleven Principal Investigators and Interdisciplinary Scientists, selected via an announcement of opportunity. Interdisciplinary Scientists are experts in specific overarching science themes connected to the mission objectives who take advantage of synergistic use of the data delivered by several experiments. Given the mission's broad scientific scope, it is standard practice for the SWT to delegate specific responsibilities to dedicated working groups (see the different groups displayed on Fig. 11). The SWT is supported by two operations groups, one focused on science operations (SOWG) and one on data archiving (DHAWG). In addition, four science-focused working groups cover moons geophysics, moon remote sensing, magnetosphere and plasma, and Jupiter's atmosphere, along with a dedicated group for analysis of the radiation monitor data. These groups contribute to scientific coordination, data analysis, trajectory assessment, and preparation of science planning. A key role of these forums is to foster scientific exchange and promote collaboration across teams, with particular emphasis on engaging early-career scientists. Two cross-disciplinary groups further support the mission: one addressing coordinate systems, cartography, and nomenclature, and another, the JUICE-Europa Clipper Steering Committee, identifying high-value joint science opportunities between the two missions. Finally, two engineering working groups currently provide support to the JUICE project in the areas of electromagnetic compatibility and radiation. In general, the possible modes of participation to the JUICE programme are: Principal Investigator, co-Principal Investigator, co-Investigator, Interdisciplinary Scientists, Guest Investigators, NASA Participating Scientists, and team associate members. The internal document "rules of the road" establishes a clear, shared framework for how people work together within the SWT and working groups. These rules define standards of professional conduct in meetings and day-to-day interactions. They aim to ensure a respectful, inclusive, and constructive working environment. Rules of the Road also explain how authorship and acknowledgements should be handled.

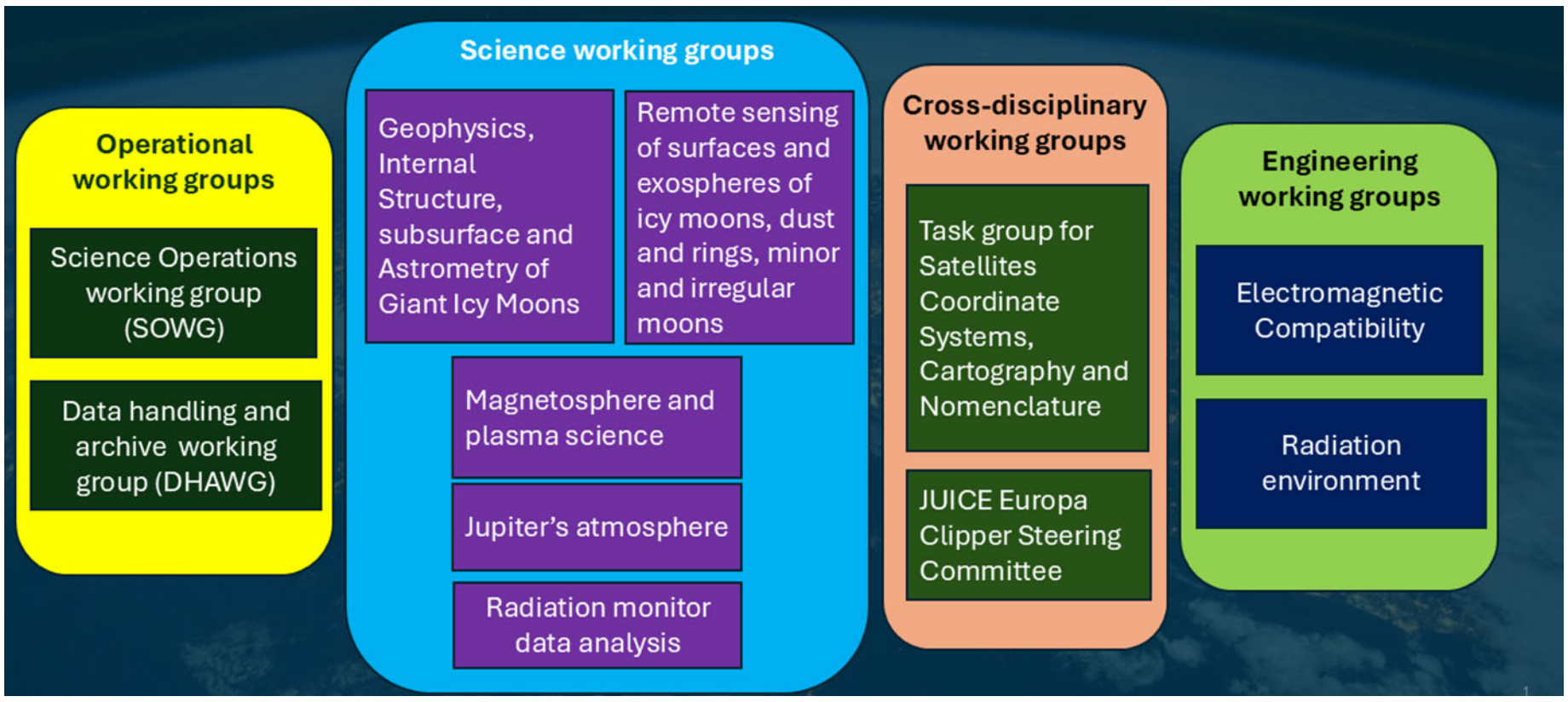

**Fig. 11** The JUICE science working team, chaired by the project scientist, is supported by various working groups as illustrated in this figure. Each working group regularly report to the SWT. Science and cross-disciplinary working groups are chaired by members of the scientific community, while operational and engineering working groups are chaired by ESA JUICE project members

## 5.3 The Launcher

Following the arrival of JUICE in French Guiana in early February 2023, there followed a two-month launch campaign and a one-day postponement due to bad weather before, on Friday, 14 April 2023, at 09:14 am local time, the spacecraft lifted off on an Ariane 5 ECA launcher (see Fig. 12). It was the 260th launch of an Ariane rocket and the penultimate Ariane 5, before the takeover of the new generation Ariane 6. The launch was perfect with all orbit insertion parameters being nominal. The Ariane 5 heavy-lift launcher was an ESA programme carried out in cooperation between public institutions and industry across 12 European countries. ArianeGroup is the lead contractor for the development and production of Ariane 5, as well as being responsible for launcher preparation operations up to lift-off. As prime contractor for Ariane 5 and the next generation Ariane 6, ArianeGroup is at the head of a vast industrial network of more than 600 companies, including 350 small and medium-size enterprises. ArianeGroup delivers a flight-ready launcher on the launch pad to its subsidiary Arianespace, which marketed and operated Ariane 5 and does the same for Ariane 6. During the launch campaign, Arianespace worked closely with the JUICE project team in charge of the spacecraft preparation and the French space agency (CNES), which is responsible for the launch base and the relevant facilities.

## 5.4 The Ground Segment

After launch and separation from the launcher, an autonomous sequence ensured that, without any need for ground intervention, the spacecraft was put in a safe and stable state with the onboard transmitter switched on, the chemical propulsions system initialised, the solar array deployed, the star tracker and all elements of the attitude and control system operative, and an attitude guaranteeing to be power positive. At this point in time the spacecraft operations began, conducted by the ground segment team. The overall JUICE ground segment is composed of the Mission Operations Centre (MOC located in ESOC) and the Science Ground Segment (SGS), which in turn is composed of the Science Operations Centre (SOC, located in ESAC) and members of the instrument's teams (see Altobelli et al. 2026, this collection). The spacecraft is operated by an 'off-line' monitoring and control approach. A pre-scheduled timeline (planned sequences of operations, defining spacecraft or instrument activities) is uploaded by the MOC at regular intervals and stored on board. Monitoring will also be off-line as the contact with ground-stations will not be continuous. All mission phases rely on ESA's 35-m deep-space ground stations, including New Norcia, Cebreros, and Malargüe. Use of the NASA Deep Space Network (DSN) is foreseen in case of critical phases and at the time of the Europa flyby phase, to support a quicker download of collected data.

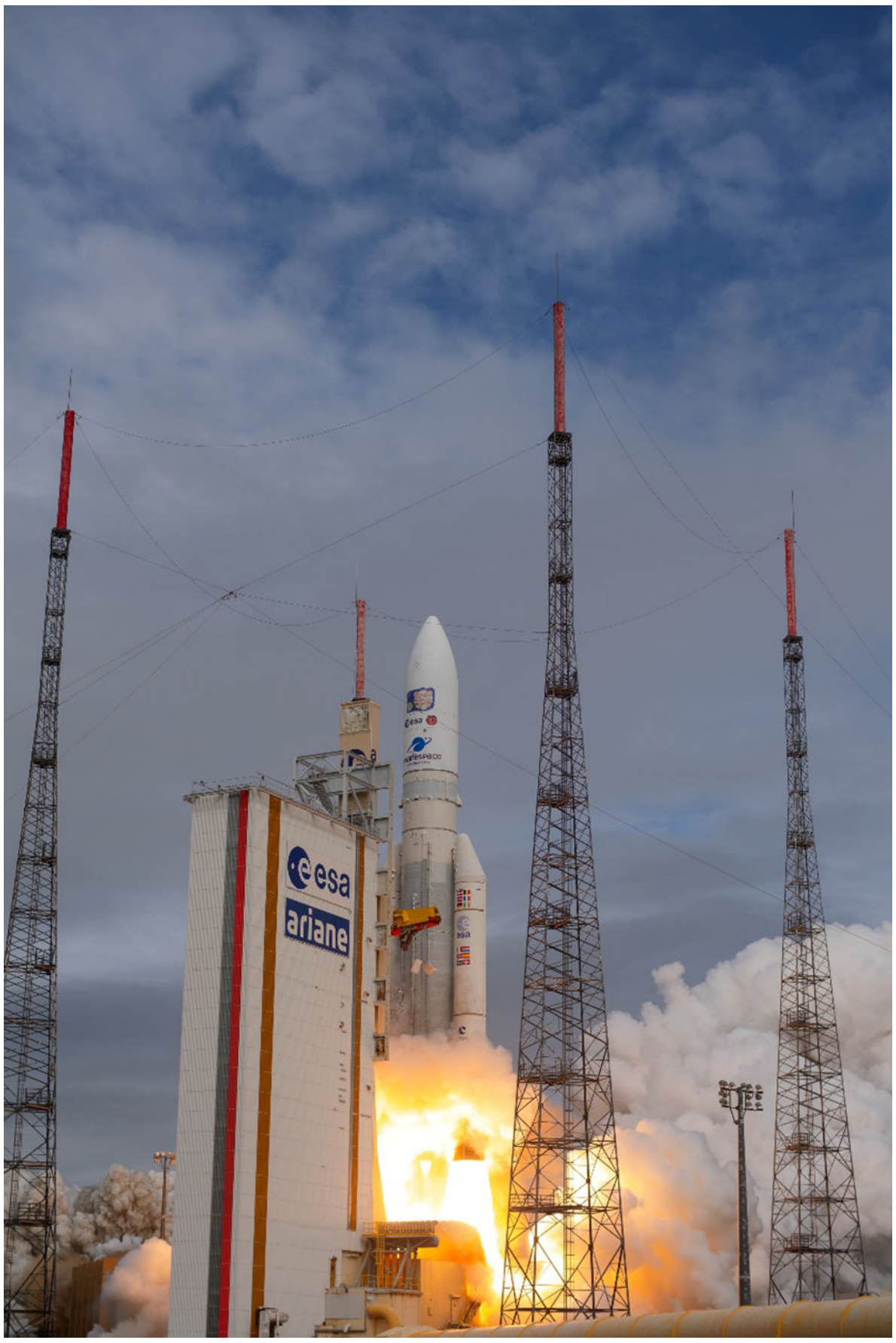

**Fig. 12** The launch of JUICE on 14 April 2023, 9:14 local time, 14:14 CEST (courtesy of ESA and Arianespace). A special piece of artwork can be seen on the fairing of the launcher. This drawing by 8-year old Yaryna was the winning entry to the 'JUICE Up Your Rocket!' competition, which invited children from all over the world to create a work of art related to JUICE

During cruise, payload checkouts are executed twice per year to check the instrument health and performance and perform calibration. Instruments are also operated during the Earth and Moon gravity assists. While it was not included in the pre-launch baseline, some science investigations are foreseen during the cruise phase, like the 3I/ATLAS campaign, a test of general relativity (di Stefano et al. 2022) and heliophysics campaigns after the third Earth flyby.

Continuous observations of the Jovian system will start a few months before arrival to Jupiter, and the insertion into Jupiter orbit will mark the start of the nominal science phase. During the nominal science operations, the ground station contact will occur daily for about 8 hours, and will be used to upload new commands, as well as to retrieve the scientific data together with the housekeeping data. No real time reaction below twelve hours is fore-seen during any mission phase on a routine basis. The science mission planning follows the standard practice of three planning cycles: long-term (six months), medium-term (few weeks) and short-term (one week). A science activity plan provides the overall strategy for the whole mission timeline.

The data, distributed to the science teams from ESAC servers, will eventually be released from the ESA planetary science archive (Besse et al. 2018), following a proprietary period of six months for the PI-led instruments. Housekeeping and ancillary data, and data from navigation cameras, monitoring cameras, radiation monitor and accelerometer will be released shortly after acquisition, processing, and sanity checks on ground. JUICE data will also be available via the NASA Planetary Data System, and a selection of data is expected to be released via other public archives such as AMDA (Génot et al. 2021).

## 6 Collaboration with Other Missions

Since its arrival at Jupiter in 2016, the Juno mission (Bolton et al. 2017) has made major discoveries and provided new insights into Jupiter's interior, atmosphere, magnetosphere, ring system, and the moons Io, Europa, and Ganymede. While these results do not modify the JUICE mission's scientific objectives, unsurprisingly given the fundamentally different goals and trajectories of the two missions, they provide very valuable inputs for future JUICE science planning and data interpretation. Juno's high-spatial resolution observations of Jupiter's deep atmosphere (via both microwave radiometry and gravitational mapping), and its detailed characterisation of Ganymede's composition, magnetosphere, ionosphere, and ice-shell structure, offer important contextual information for the future JUICE observations. In particular, following a single very successful flyby of Ganymede in 2021, the Juno observations provided valuable information on the surface chemistry (salts, organics), the magnetosphere coupling and reconnection, and the aurora and plasma physics (e.g. Ravine et al. 2022; Molyneux et al. 2022; Hue et al. 2023; Tosi et al. 2024b; Benmahi et al. 2025; Moirano et al. 2026; Stahl et al. 2023; Joseph et al. 2024). Likewise, Juno's latest measurements of Jupiter's plasma, and magnetic environment, as well as updated constraints on the surrounding space environment, will support internal reassessments of the Jovian radiation conditions relevant for JUICE performances and operations.

Europa Clipper (Pappalardo et al. 2024), launched on 14 October 2024 from Cape Canaveral on a Falcon Heavy 9, will reach Jupiter about one year before JUICE due to the more powerful rocket, and spend a year in the Jovian system before conducting the planned 49 Europa flybys during its nominal three-year mission, along with serendipitous encounters with Ganymede and Callisto. The simultaneous presence of two highly instrumented spacecraft in the Jovian system offers unprecedented opportunities for synergistic observations during their orbital phases, as well as unique heliophysics and magnetospheric studies during cruise phase and Jupiter approach. Although no formal commitments currently exist for coordinated science beyond each mission's primary objectives, collaboration between the two teams began during spacecraft development, and joint workshops, conference sessions, and a dedicated JUICE–Europa Clipper Steering Committee has helped identify high-value joint science opportunities through joint trajectory analysis. Potential cruise-phase activities include studying solar wind plasma, interplanetary magnetic fields, and solar wind structure when the spacecraft are aligned in heliospheric longitude, radially aligned, at similar heliocentric distances, or magnetically connected via Parker spiral field lines. Another example, both JUICE-UVS and Europa-UVS (Retherford et al. 2024) simultaneously observed interstellar comet 3I/ATLAS.

In particular, during JUICE's Jupiter approach while Europa Clipper orbits inside the magnetosphere, the missions jointly can investigate solar wind-magnetosphere interactions, auroral processes, and the influence of solar wind variability on the moons' local plasma environments. After JUICE Jupiter orbit insertion in July 2031, several coordinated opportunities arise throughout the Jovian system, including during JUICE's second Europa flyby (16 July 2032) that coincides with the Europa Clipper flyby E23. During this opportunity the two spacecraft will fly past Europa within 0.5 $R_J$ (Jupiter radii) of each other and just a few hours apart and will remain within 2 $R_J$ of each other for approximately 2 weeks. Such opportunities allow contemporaneous or asynchronous measurements, cross-calibration efforts, and serendipitous science. A consolidated set of joint objectives is expected to be documented in a future publication. Finally, because both spacecraft are expected to end their missions by impacting Ganymede, one may be able to observe the other's impact, including any ejecta plume and the resulting crater.

The cruise-phase science return from JUICE (and Europa Clipper) could be further strengthened by enabling multi-point measurements of space-weather events across the heliosphere by incorporating data from other active spacecraft, such as BepiColombo at Mercury (Benkhoff et al. 2021), Solar Orbiter between 0.284 AU and about 1 AU (Mueller et al. 2020), Parker Solar Probe between 0.046 AU and about 0.7 AU (Fox et al. 2016), PUNCH in Earth orbit (DeForest et al. 2026), IMAP at L1 (Hegarty et al. 2026), and Psyche between 1 AU and 3.1 AU (Polanskey et al. 2025), thereby broadening the range of interesting geometric configurations (e.g., Parker spiral connections; Turner and Li 2011).

If resolved by its imaging instruments, comparison of observations of the small and irregular moons of Jupiter by JUICE with observations of the Jovian Trojans by NASA's Lucy (Levison et al. 2025) will provide important clues on the degree of parenthood and taxonomic similarities between Jovian small moons and Trojans.

## 7 Coordination with Earth-Based Observatories

Coordination between Solar System exploration missions and ground-based observatories has consistently been proven to be highly effective (e.g., Orton et al. 1998; Witasse et al. 2006; Lellouch and Witasse 2008; Nichols et al. 2017; Palmaerts et al. 2024). In line with these past efforts, ground-based observations will be an essential addition to the JUICE mission, providing contextual information for JUICE measurements and offering complementary spatial, temporal, and spectral coverage beyond the payload wavelength range (Fletcher et al. 2023). Observations in the X-ray, EUV, mid- and far-infrared, and radio domains will be particularly valuable. Observations by ground-based radio telescopes of the JUICE communication signal are already incorporated through the PRIDE experiment (Gurvits et al. 2023, this collection). Ground-based support is expected to address several key areas, including Jupiter's atmospheric dynamics and composition, auroral activity, the surfaces and exospheres of the Galilean moons, Ganymede's auroras, astrometry, and ephemerides (particularly via stellar occultations), Io's volcanic activity, and Jupiter's radio emissions. Coordination with ground- or space-based assets during some of the JUICE-Europa Clipper joint measurements can also increase the scientific return of both missions.

Coordination is not limited to ground-based observatories – both the Hubble Space Telescope and the James Webb Space Telescope (JWST) are expected to be operating during the JUICE mission in the early 2030s, providing high-resolution global imaging in the visible, near, and mid-IR, alongside spectroscopic maps at a higher spectral resolution (but lower spatial resolution) than can be achieved from JUICE. Complementary observations of the atmosphere, ionosphere, satellites and rings could be planned at important moments within the JUICE tour and Ganymede orbits.

## 8 Concluding Remarks

JUICE's future studies of the Jupiter system will enhance our understanding of the evolution and dynamics of giant planets and their moons. The mission is built around two overarching science themes: the emergence of habitable worlds around gas giants and the Jupiter system as an archetype for gaseous giant planets. JUICE's diverse, wide-ranging, and interdisciplinary science objectives will be pursued through close collaborations among the science teams, with the aim of maximizing overall the largest scientific return. Collaborations with the Europa Clipper teams and ground-based observers are expected to be particularly useful and constructive.

As illustrated in the timeline presented in Table 1, the study and development phases of the programme were both extensive and highly complex, involving numerous significant challenges. On the technical side, the main difficulty was to design and develop a large space probe that will navigate mostly autonomously in the harsh environment of a distant planetary system. The programmatic issues had to do with keeping the schedule and the cost of the mission within the required boundaries agreed at the time of the adoption of the mission in 2014, but also coordinating a large and internationally distributed industrial and scientific consortium, and holding partnership with members states' Space Agencies as displayed on Fig. 13. Dealing with the possibility for a launcher change from Ariane 5 (the baseline launcher) to Ariane 6 (the back-up launcher, but still under development at that time) also introduced an element of uncertainty and schedule pressure (Sarri et al., this collection). Finally, two challenges emerged in the final years of the development: the Covid-19 pandemic and the onset of the Russian invasion of Ukraine. Dealing with Covid in the years 2020–2022 required a new mindset and ways of working. While the teleworking was becoming the standard, this was not possible for the spacecraft integration and tests that needed to be performed physically on site. A very complex organisation had to be put in place which was very successful in the face of unprecedented difficulties. The war in Eastern Europe had a direct impact on the availability of the Antonov-124 air freight service normally used to transport most of the European spacecraft, and therefore JUICE, from Europe to Kourou. The impact on JUICE was limited to the uncertainty on the transport, which had to be dealt with and luckily worked out well.

Following its launch on 14 April 2023, the JUICE spacecraft underwent a three-month period of meticulous in-orbit verification, known as commissioning, which concluded successfully on 19 July 2023 with an ESA high-level review. Experts and managers confirmed that all systems and instruments were operating nominally and that the development phase of the mission had been successfully completed. After performing a few deep-space manoeuvres and carrying out the Moon–Earth (Dietz et al. 2026; Vallat et al. 2026) and Venus's flybys, the spacecraft is now *en route* to perform its second Earth gravity assist and, at the time of writing, has covered more than one third of its journey toward its destination: Jupiter and its icy moons. The cruise phase is largely dedicated to navigation, regular checks of both payload and spacecraft, essential instrument calibrations, software updates, and preparations for the science phase and the future data archive. Nevertheless, scientifically valuable campaigns are being organised around the gravity assists and during special opportunities, such as the encounter with 3I/ATLAS, for the benefit of the scientific community. Figure 14 displays a selection of in-flight observations and measurements.

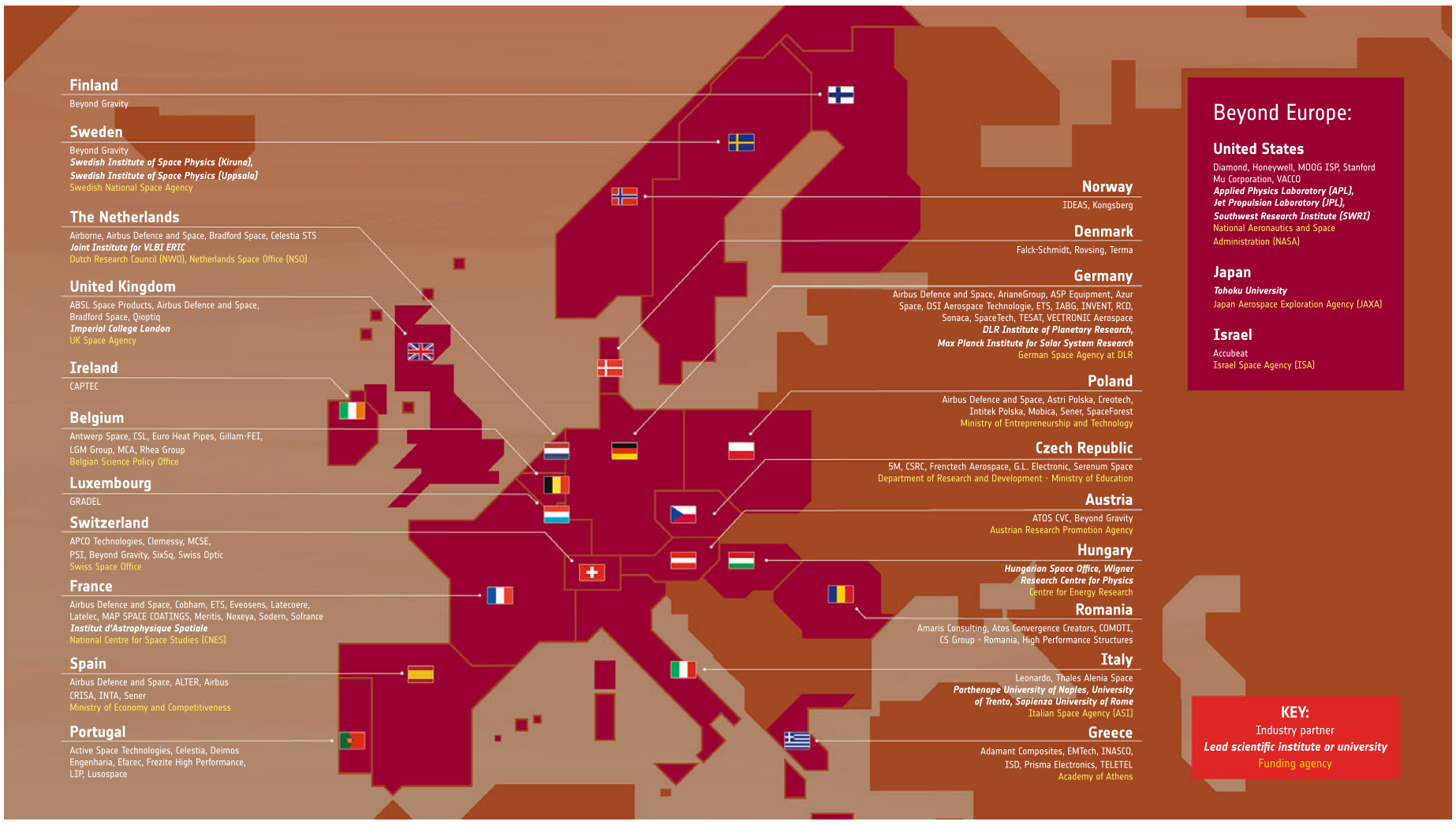

**Fig. 13** JUICE has been developed and built with contributions from many ESA Member States. This map highlights industry partners and the lead scientific institutes and universities that have contributed to JUICE. Participating countries outside of Europe are also indicated (courtesy of ESA)

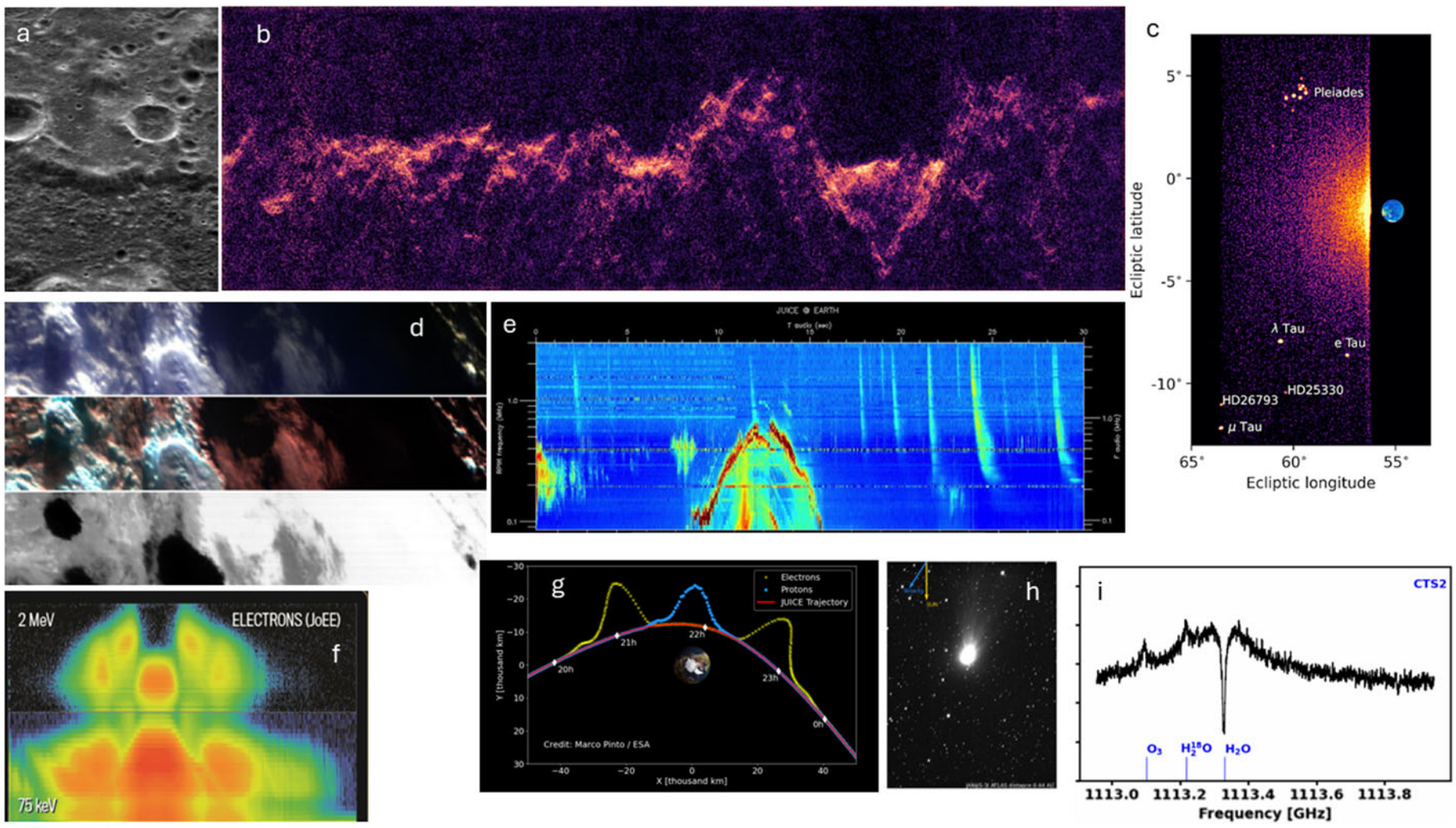

**Fig. 14** Selection of in-flight observations and measurements. a) JANUS image of the Moon taken during the Moon flyby in August 2024. ESA/JUICE/JANUS. b) RIME radargram acquired during the Moon flyby in August 2024. Credits: ESA/JUICE/RIME. c) Geocoronal hydrogen atoms radiating from the Earth recorded by the UVS instrument. Credit Southwest Research Institute. d) Three horizontal strips showing different images of Earth cloudy structures by MAJIS taken in a different wavelength, during the Earth flyby in August 2024. Credits: ESA/JUICE/MAJIS. e) Radio waves recorded by RPWI during the Earth flyby in August 2024. Credits: ESA/JUICE/RPWI. f) Electron fluxes measured by PEP during the Earth flyby in August 2024. Credits: ESA/JUICE/PEP. g) Earth radiation belts measured by RADEM. Credits: ESA/JUICE/RADEM. h) 3I/ATLAS observed by JANUS in November 2025. Credits: ESA/JUICE/JANUS. i) Atmospheric composition measured by SWI during the Earth flyby. Credits: ESA/JUICE/SWI

JUICE was identified and selected in 2012 as the natural next step in the exploration of the outer Solar System, building on the legacy of the Galileo (e.g. Bagenal et al. 2004), Cassini–Huygens (e.g. Dougherty et al. 2009) and Juno (e.g. Bolton et al. 2017) missions. And the journey will not end there. Following JUICE, the European Space Agency plans to push even farther with 'L4,' the first large mission of the Voyage 2050 programme, which will travel to the Saturnian system, explore the icy moon Enceladus, and investigate other ocean worlds.

**Acknowledgements** JUICE is a mission under ESA leadership with contributions from its Member States, NASA, JAXA and the Israel Space Agency. It is the first Large-class mission in ESA's Cosmic Vision programme. The mission represents the culmination of many years of dedicated effort, technical excellence and sustained commitment by numerous teams and institutions. The authors wish to acknowledge the outstanding work of the staff at ESA, whose leadership, coordination and expertise have been fundamental to the development and successful implementation of the mission. We further recognise the major contribution of European industry, led by Airbus Defence and Space, whose engineering excellence and system-level expertise have enabled the realisation of this highly complex spacecraft. We gratefully acknowledge the scientific community, led by the Principal Investigators of the JUICE instruments, whose vision and scientific leadership have shaped the mission's objectives and payload. We thank the national funding agencies and supporting institutions whose continuous support has made this ambitious programme possible. OW thanks Arvind Parmar for his trust. The documentary 'The making of JUICE', looking behind the scenes at the development and testing of the spacecraft in the three years leading up to launch, is available on YouTube. JUICE webinars organised by Europlanet are also available on Youtube. A selection of JUICE data is already available on the ESA planetary science archive https://psa.esa.int. The launch media kit: https://esamultimedia.esa.int/docs/science/JUICE-LaunchKit.pdf. Winners of the fairing drawing competition: https://www.esa.int/kids/en/news/Children_of_the_world_join_Europe_s_mission_to_Jupiter. 3I/ATLAS preliminary findings can be found at: https://www.esa.int/Enabling_Support/Operations/A_rare_encounter_How_Juice_came_to_observe_3I_ATLAS. The mission can be written JUICE as an acronym or Juice, in accordance with ESA guidelines for style consistency between space missions when communicating with the public.

**Funding Information** Open Access funding enabled and organized by Projekt DEAL.

## Declarations

**Competing Interests** The authors declare that they have no competing interests. OW is the Editor-in-Chief of Space Science Reviews, but was not involved in the peer-review process of this article.

## Authors and Affiliations

**Olivier Witasse[1] · Giuseppe Sarri[1] · Christian Erd[1] · Philippe Gare[1] · Nicolas Altobelli[2] · Claire Vallat[2] · Santa Martinez[2] · Rosario Lorente[2] · Angela Dietz[3] · Arnaud Boutonnet[3] · Frank Budnik[3] · Ignacio Tanco[3] · Cyril Cavel[4] · Frederic Faye[4] · Didier Radola[4] · Vincent Poinsignon[4] · Pasquale Palumbo[5] · Francois Poulet[6] · Giuseppe Piccioni[5] · Yves Langevin[6] · Kurt D. Retherford[7,8] · Randy Gladstone[7] · Paul Hartogh[9] · Luciano Iess[10] · Hauke Hussmann[11] · Giuseppe Cimò[12,13] · Leonid I. Gurvits[13] · Lorenzo Bruzzone[14] · Michele K. Dougherty[15] · Stas Barabash[16] · Jan-Erik Wahlund[17] · Olivier Grasset[18] · Leigh N. Fletcher[19] · Norbert Krupp[9] · Gabriele Cremonese[20] · Ganna Portyankina[11] · Yohai Kaspi[21] · David Stevenson[22] · Dominik Dirkx[13] · Jeffrey J. Plaut[23] · Pontus Brandt[24] · Peter Wurz[25] · Baptiste Cecconi[26] · Ingo Mueller-Wodarg[15] · Hanna Rothkaehl[27] · Ondrej Santolík[28] · Patricia Gonçalves[29] · Wojtek Hajdas[30] · Emma J. Bunce[19] · Tim Van Hoolst[31,32] · Gabriel Tobie[18] · Thomas Roatsch[11] · Federico Tosi[5] · Cecilia Tubiana[5] · John Carter[6] · Adam Masters[15] · Ronan Modolo[33] · Elias Roussos[9] · Thibault Cavalié[34] · Alexander Stark[11] · Elke Kersten[11] · Tilmann Denk[11] · Athena Coustenis[26] · Dmitrij Titov[35] · Michel Blanc[36,37] · Jean-Pierre Lebreton[38] · Veronique Loisel[39]**

✉ Olivier Witasse
Olivier.Witasse@esa.int

✉ Paul Hartogh
hartogh@mps.mpg.de

✉ Norbert Krupp
krupp@mps.mpg.de

1 European Space Agency, ESTEC, Keplerlaan 1, Noordwijk, The Netherlands

2 European Space Agency, ESAC, Camino Bajo del Castillo, s/n, 28692 Villafranca del Castillo, Madrid, Spain

3 European Space Agency, ESOC, Robert-Bosch-Straße 5, Darmstadt, Germany

4 Airbus Defence and Space SAS, 31 Rue des Cosmonautes, Toulouse, France

5 Istituto Nazionale di Astrofisica—Istituto di Astrofisica e Planetologia Spaziali (INAF-IAPS), Rome, Italy

[6] Institut d'Astrophysique Spatiale, CNRS/Université Paris-Saclay, 91405 Orsay Cedex, France

[7] Southwest Research Institute, 6220 Culebra Road, San Antonio, TX 78238, USA

[8] University of Texas at San Antonio, 1 UTSA Circle, San Antonio, TX 78249, USA

[9] Max-Planck-Institut für Sonnensystemforschung, Göttingen, Germany

[10] Dipartimento di ingegneria meccanica e aerospaziale, Centro Ricerca Aerospaziale Sapienza (CRAS), Università La Sapienza, via Eudossiana 18, 00184 Rome, Italy

[11] Institute of Space Research, DLR, Deutsches Zentrum für Luft- und Raumfahrt, Rutherfordstr. 2, 12489 Berlin, Germany

[12] Joint Institute for VLBI ERIC, Oude Hoogeveensedijk 4, 7991 PD, Dwingeloo, The Netherlands

[13] Delft University of Technology, Kluyverweg 1, 2629 HS, Delft, The Netherlands

[14] Department of Information Engineering and Computer Science, University of Trento, Via Sommarive, 5, I-38123, Trento, Italy

[15] Department of Physics, Imperial College London, London SW7 2AZ, UK

[16] Swedish Institute of Space Physics (IRF), Box 812, 981 28 Kiruna, Sweden

[17] Swedish Institute of Space Physics (IRF), Box 537, 751 21 Uppsala, Sweden

[18] Laboratoire de Planétologie et Géosciences, UMR-CNRS 6112, Nantes Université, 2, rue de la Houssinière, 44322 Nantes cedex, France

[19] School of Physics and Astronomy, University of Leicester, University Road, Leicester, LE1 7RH, UK

[20] INAF - OAPd (Osservatorio Astronomico di Padova), Vicolo Osservatorio 5, 35122, Padova, Italy

[21] Department of Earth and Planetary Sciences, Weizmann Institute of Science, Rehovot, Israel

[22] Division of Geological and Planetary Sciences, California Institute of Technology, Pasadena, CA 91125, USA

[23] Jet Propulsion Laboratory, 4800 Oak Grove Drive, Pasadena, CA 91109, USA

[24] Applied Physics Laboratory, Johns Hopkins University, 11100 Johns Hopkins Rd, Laurel, MD 20723, USA

[25] Space Sciences and Planetology, Physics Institute, University of Bern, Sidlerstrasse 5, 3012, Bern, Switzerland

[26] LIRA, Observatoire de Paris, Université PSL, Sorbonne Université, Université Paris Cité, CY Cergy Paris Université, CNRS, 92190 Meudon, France

[27] Space Research Centre of the Polish Academy of Sciences, Bartycka 18A, 00-716 Warszawa, Poland

[28] Institute of Atmospheric Physics of the Czech Academy of Sciences, Boční II 1401, Prague 4, Czech Republic

[29] Laboratory for Instrumentation and Experimental Particle Physics, Av. Prof. Gama Pinto 2, Lisbon, 1649-003, Portugal

[30] Paul Scherrer Institut, Forschungstrasse 111, Villigen PSI, 5232, AG, Switzerland

[31] Royal Observatory of Belgium, Ringlaan 3, 1180 Brussels, Belgium

[32] Institute of Astronomy, KU Leuven, Celestijnenlaan 200D, 3001 Leuven, Belgium

[33] LATMOS, Sorbonne University, 4 Pl. Jussieu, 75005 Paris, France

[34] Laboratoire d'Astrophysique de Bordeaux, Univ. Bordeaux, CNRS, B18N, allée Geoffroy Saint-Hilaire, 33615 Pessac, France

[35] Observatory of Leiden, Gorlaeus Building, Einsteinweg 55, 2333 CC, Leiden, The Netherlands

[36] IRAP, 9, avenue du Colonel Roche, BP 44346, 31028 Toulouse, France

[37] Weihai Institute of Interdisciplinary Research, Shandong University, Weihai, China

[38] CNRS-LPC2E, Université d'Orléans, 3A, Avenue de la Recherche Scientifique, 45071 Orléans cedex, France

[39] Arianespace SAS, C/O ArianeGroup, 51-61 Route de Verneuil, 78130 Les Mureaux, France